\documentclass[usegraphicx]{emulateapj} 
\usepackage{natbib}

\newcommand{\sm}{\sim\!}
\newcommand{\keVpar}{\,keV\,particle$^{-1}$}
\newcommand{\keVcm}{\,keV\,cm$^{2}$}
\newcommand{\ergss}{\,ergs\,s$^{-1}$}
\newcommand{\msun}{\,M$_\odot$}
\newcommand{\rvir}{r_{\mathrm{vir}}}
\newcommand{\rc}{r_{\mathrm{c}}}
\newcommand{\lx}{L_{\mathrm X}}
\newcommand{\tx}{T_{\mathrm X}}
\newcommand{\kx}{K_{\mathrm X}}
\newcommand{\mx}{M_{\mathrm X}}
\newcommand{\sx}{\sigma_{\mathrm X}}
\newcommand{\stx}{T^\star_{\mathrm X}}
\newcommand{\stg}{T^\star_{\mathrm g}}
\newcommand{\DE}{\Delta E}
\newcommand{\sDE}{\DE^\star}
\newcommand{\tvir}{T_{\mathrm{vir}}}
\newcommand{\stvir}{T^\star_{\mathrm{vir}}}
\newcommand{\tstg}{\widetilde{T}^\star_{\mathrm g}}
\newcommand{\Dvir}{\Delta_{\mathrm{vir}}}
\newcommand{\Dthr}{\Delta_{\mathrm{m}}}
\newcommand{\rhou}{\rho_{\mathrm u}}
\newcommand{\rhoc}{\rho_{\mathrm c}}
\newcommand{\rhox}{\rho_{\mathrm X}}
\newcommand{\rhog}{\rho_{\mathrm g}}
\newcommand{\rhogc}{\rho_{\mathrm{g,c}}}
\newcommand{\trhog}{\widetilde{\rho}_{\mathrm g}}
\newcommand{\trhogiso}{\widetilde{\rho}_\mathrm{g,iso}}
\newcommand{\nx}{n_{\mathrm X}}
\newcommand{\rs}{r_{\mathrm s}}
\newcommand{\zf}{z_{\mathrm{for}}}
\newcommand{\Om}{\Omega_{\mathrm m}}
\newcommand{\Omb}{\Omega_{\mathrm b}}
\newcommand{\Mc}{M_{\mathrm c}}
\newcommand{\Vc}{V_{\mathrm c}}

\newcommand{\Phic}{\Phi_{\mathrm c}}
\newcommand{\mub}{\bar{\mu}}
\newcommand{\mubpre}{\mub_{\mathrm{pre}}}
\newcommand{\msim}{M_{\mathrm{500}}}
\newcommand{\csim}{c_{\mathrm{500}}}
\newcommand{\msimfin}{M_{\mathrm{500,13}}}
\newcommand{\mxsim}{M_{\mathrm{X,500}}}
\newcommand{\txsim}{T_{\mathrm{X,500}}}
\newcommand{\fxsim}{F_{\mathrm{X,500}}}
\newcommand{\ctwohun}{c_{\mathrm{200}}}
\newcommand{\coneight}{c_{\mathrm{180}}}

\shorttitle{X-Ray Properties of Nearby Galaxy Systems}
\shortauthors{Solanes et al.}

\begin{document}

\title{Implications of Halo Inside-out Growth on the X-Ray Properties of Nearby Galaxy Systems within the Preheating Scenario}


\author{Solanes, J.M.}
\affil{Departament d'Astronomia i
Meteorologia\altaffilmark{1}, Universitat de Barcelona, Av.\ Diagonal 647, 08028~Bar\-ce\-lo\-na, Spain; jm.solanes@ub.edu}
 
\author{Manrique, A.}
\affil{Departament d'Astronomia i
Meteorologia, Universitat de Barcelona, Av.\ Diagonal 647, 08028~Bar\-ce\-lo\-na, Spain; a.manrique@ub.edu}

\author{Gonz\'alez-Casado, G.}
\affil{Departament de Matem\`atica Aplicada II, Universitat Polit\`ecnica de Catalunya, Edificio Omega, Campus Nord, Jordi Girona 1--3, 08034~Bar\-ce\-lo\-na, Spain; guillermo.gonzalez@upc.edu}

\and

\author{Salvador-Sol\'e, E.}
\affil{Departament d'Astronomia i
Meteorologia\altaffilmark{1}, Universitat de Barcelona, Av.\ Diagonal 647, 08028~Bar\-ce\-lo\-na, Spain; e.salvador@ub.edu}

\altaffiltext{1}{Also at the Centre Especial de Recerca en Astrof\'\i
sica, F\'\i sica de Part\'\i cules i Cosmologia associated with the
Instituto de Ciencias del Espacio, Consejo Superior de Investigaciones
Cient\'\i ficas.}

             
\begin{abstract} 

We present an entirely analytic model for a preheated, polytropic
intergalactic medium in hydrostatic equilibrium within a NFW dark halo
potential in which the evolution of the halo structure between major
merger events proceeds inside-out by accretion. This model is used to
explain, within a standard $\Lambda$CDM cosmogony, the observed X-ray
properties of nearby relaxed, non-cooling flow groups and clusters of
galaxies. We find that our preferred solution to the equilibrium
equations produces scaling relations in excellent agreement with
observations, while simultaneously accounting for the typical
structural characteristics of the distribution of the diffuse
baryons. In the class of preheating models, ours stands out because it
offers a unified description of the intrahalo medium for galaxy systems
with total masses above $\sm 2\times 10^{13}$\msun, does not produce
baryonic configurations with large isentropic cores, and reproduces
faithfully the observed behavior of the gas entropy at large radii. All
this is achieved with a moderate level of energy injection of
about half a keV, which can be easily accommodated within the limits of
the total energy released by the most commonly invoked feedback
mechanisms, as well as with a polytropic index of 1.2, consistent with
both many observational determinations and predictions from
high-resolution gas-dynamical simulations of non-cooling flow
clusters. More interestingly, our scheme offers a physical motivation
for the adoption of this specific value of the polytropic index, as it
is the one that best ensures the conservation after halo virialization
of the balance between the total specific energies of the gas
and dark matter components for the full range of masses investigated.

\end{abstract}


\section{INTRODUCTION}\label{intro}

In this paper we explore the consequences of applying a recently
developed procedure for the evolution of the halo structure onto the
modeling of the properties of the hot gaseous baryons they host. Our
model assumes, in line with the results of numerical simulations
\citetext{see \citealt*{SMS05}}, that between major mergers halos grow
inside-out maintaining invariant their structural parameters. This
concept has been the basis for the successful reproduction of both the
NFW \citep*{NFW97} functional form of the dark matter density
distribution and the empirical correlations involving its concentration
parameter \citep{Man03,SMS05}, while it has also been invoked to
resolve some apparent challenges to the CDM model for structure
formation \citep{LP03}. We are interested in confronting the
predictions of this halo growth process with both the observed global
and structural X-ray properties of nearby groups and clusters of
galaxies and, in particular, in investigating its compatibility with
the findings of both observations and high-resolution gas-dynamical
simulations that the diffuse baryons outside the innermost regions of
galaxy systems obey a polytropic equation of state that is neither
purely isothermal nor adiabatic.

It has been known for nearly two decades that X-ray galaxy systems are
at variance with the simple self-similar scenario predicted by
\citet{Kai86} in which gravitational collapse is the sole mechanism
driving the evolution of their nonbaryonic and baryonic
components. Provided that galaxy groups and clusters of total virial
mass $M$ can be treated as scaled versions of each other and that their
X-ray emitting gas is in hydrostatic equilibrium within the dark matter
(DM) potential, this scenario assumes that the virial
temperature\footnote{In the present work, temperatures expressed in
\emph{specific} energy units will be identified by a star superscript.}
of the dark halo, $\stvir=GM/2\rvir$, is a good approximation for the
\emph{effective X-ray gas temperature}, $\stx\equiv k\tx/\mub=\sx^2$,
where $\rvir$ is the virial radius, $\sx$ the 1D velocity dispersion of
the gas, and $\mub=\mu m_{\mathrm p}$ the mean molecular weight of the
gas in physical units, $\mu$ being the corresponding value in units of
the proton mass $m_{\mathrm p}$ and $k$ the Boltzmann's constant. Under
these conditions it readily follows that $\tx\propto M^{2/3}$ and, if
the total X-ray gas mass $\mx\propto M$, that $\tx\propto\mx^{2/3}$
too. Furthermore, if the plasma emissivity is dominated by thermal
bremsstrahlung ---this is strictly true only for the hotter,
non-relativistic plasmas with $\tx\gtrsim 3\ \mathrm{keV}\simeq 3\times
10^7\ \mathrm{K}$, as at lower temperatures, emission line cooling by
metals dominates---, its integral over the entire energy range of the
(continuum) X-ray emission and over the gas distribution, assumed to be
an isothermal plasma with (number) density $\nx\equiv\rhox/\mub$
proportional to the mass density distribution and extending likewise
out to $\rvir$, predicts a total halo bolometric X-ray luminosity
$\lx\propto\tx^2$ or, equivalently, $\lx\propto M^{4/3}$.

While hydrodynamic simulations dealing only with physical processes
that do not have any preferred scale
\citep*[e.g.,][]{NFW95,EMN96,BN98,ENF98,Fre99,Tho01} have extensively
confirmed the approximate validity of the self-similar scaling even for
cosmogonies of the CDM type, years of observational efforts have now
provided compelling enough evidence for an excess in the specific
energy/entropy of the X-ray gas with respect to the contributions
arising from adiabatic compression and shock heating alone
\citep*[e.g.,][]{Whi91,DFJ91,DJF96,PCN99,LPC00,Fin02,PSF03}. These
deviations are indicative of an important role of nongravitational
processes in determining the properties of the hot intrahalo medium
(hereafter IHM).

Basically, there are two different approaches competing to explain the
observed X-ray properties of bound collections of galaxies and, in
particular, the deviations from the simple self-similar picture. One
category of models (both analytical and numerical) explores the effects
of nongravitational heating on the IHM drawing inspiration from the
work of \citet{Kai91}, who was the first to point out that the
intracluster gas properties, including the evolution of its X-ray
luminosity function, could be better reconciled with the hierarchical
scenario of structure formation if the entropy of the gas we now see
was the result of some early injection of heat ---perhaps during the
epoch of galaxy formation--- previous to the cluster assembly. The
addition of energy into the gaseous component increases its entropy and
reduces the shock-heating efficiency, preventing the gas from reaching
high density during the halo collapse and producing lower than expected
luminosities for a given temperature
\citep*[e.g.,][]{CMT98,WFN98,BBP99,Loe00,WFN00,TN01,
BEM01,Bor01,Bab02,MBB02,DD02,OB03}. For a given amount of energy
$\DE$ injected into the intergalactic medium, this extra heating
is expected to leave a more significant imprint on the low-temperature
systems for which $\DE/E\gtrsim 1$ and the shock-created entropy
is negligible, in agreement with observations. This means that for
these systems gas infall proceeds roughly adiabatically, predicting the
existence of an 'entropy floor' \citep{LPC00}. Paradoxically, radiative
cooling is also capable of explaining the lack of self-similarity of
the X-ray gas \citep*[e.g.,][]{Bry00,VB01,Mua01,WX02,DKW02}. In this
alternative scenario, the central low entropy gas, characterized by a
cooling time shorter than the typical halo age, is selectively removed
by condensing into dense, cold structures and rapidly replaced by the
higher-entropy material from the halo outskirts. The latter is then
heated by adiabatic compression as it flows in, giving rise to
temperature profiles with a central maximum and engendering
substantially shallower density distributions that cause a reduction
in the bulk X-ray luminosity, much like in the non-gravitational
heating scheme.

These two basic scenarios, however, are not free of important
drawbacks. Radiative cooling without non-gravitational feedback
mechanisms is a runaway process that results in severe overcooling of
gas on group scales and thereby in an unpleasantly large fraction of
gas ($\sm 50\%$) converted into a 'stellar' cold medium
\citep{OB03}. This so-called 'cooling crisis' or 'cooling catastrophe'
\citep*[e.g.,][]{BVM92,FHP98,Bal01} is difficult to be dealt with
appropriately even with state-of-the-art numerical simulations due to
the technical challenge represented by the simultaneous requirement of
properly implementing small-scale processes, such as star formation and
feedback, and having a large dynamic range from galaxies to large-scale
structures in a volume large enough to guarantee a statistically
representative ensemble of simulated halos. Overcoming overcooling is
not easy either within analytical radiative cooling schemes
\citep[e.g.,][]{Voi02}. Important cooling and condensation of the gas,
and the internal heating resulting from the ensuing star formation
feedback, continuously modify the gas distribution, so the necessary
hypothesis that the IHM is in hydrostatic equilibrium does not
hold. Besides, any heating of the intergalactic medium after it is
confined within halos of groups and clusters has the disadvantage of
requiring a larger amount of energy injection to raise the denser gas
to a given entropy level \citep{Loe00,VB01}. Precisely, the major
criticism toward the preheating scenario has to do with the 'energy
crisis' \citep[e.g.,][]{DKW02} related to the yet unidentified
astrophysical source(s) responsible for injecting the amount of excess
energy ($\sm 1$\keVpar) necessary to explain the observed scaling
relations and its connection with the process of galaxy formation
\citep[e.g.,][]{NFW95,CMT98,BBP99,LPC00,WFN00,BM01,Bab02}. We want to
point out, however, that although important details on how and when
this additional heat is injected into the diffuse medium have still to
be sorted out, there is no shortage of potential sources of
pre-collapse heating \citetext{see, for instance, \citealt{Bab02} for a
brief summary}. More importantly, the substantial additional entropy at
large radii detected in both clusters and groups of galaxies at a
typical level of $\sm 400$\keVcm\ \citep*{Fin02,PSF03} appears to be
beyond the bounds of plausibility of models that invoke only gas
cooling and gravitational heating to explain the $\sm
100\,h^{-1/3}$\keVcm\ entropy excess seen in the central regions of
galaxy systems \citep{PCN99,LPC00,PSF03}. Two recent results give
convincing support to this conclusion. To begin with, there are the
hydrodynamic simulations by \citet{Fin03} with radiative cooling, star
formation, and non-gravitational heating, which demonstrate that star
formation without extra heating produces too steep entropy profiles in
group-sized halos that also fall somewhat short in explaining the
height of the observed specific entropy on the halo outskirts. In
addition, observations of the hot cluster Abell 1795 by \citet*{IBK04}
show that the intracluster medium temperature is higher than the dark
matter 'temperature' (measured from the radial profile of the dark
matter velocity dispersion), even in the central region where the
radiative cooling time is short.

While it is true that recent high spatial resolution X-ray data suggest
that treatments combining energy injection with cooling may be
necessary to explain the central temperature and entropy profiles of
massive cooling flow clusters \citep[e.g.,][]{McC04}, this work aims to
demonstrate that a hydrostatic polytropic gas model based on the notion
of preheating and requiring a relatively modest heating energy budget
can have a remarkable success in accounting simultaneously for the
global and structural X-ray properties \emph{representative of the bulk
of present-day relaxed galaxy systems} with gravitational masses
between $\sm 10^{13}$ and $\sm 10^{16}$\msun. This is done in the
context of the concordance flat $\Lambda$CDM cosmology with reduced
Hubble constant $h=2/3$, present-day matter density $\Om=1/3$, rms mass
fluctuation on scales of 8 $h^{-1}$ Mpc $\sigma_8=0.95$, cosmological
baryon content $\Omb=0.04$ (and, hence, a cosmic baryon fraction
of 0.12), and primordial mass density fluctuation power spectrum index
$n=1$, fully consistent with recent joint analyses of \emph{WMAP} and
redshift survey data \citep[e.g.,][]{Teg04}. The outline of the paper
is as follows. In Section~2, we provide a brief overview of the
approach used to follow the growth of bound halos and present the
profiles describing the structure and kinematics of their nonbaryonic
and baryonic components. After identifying the best solution to the
equilibrium equations of the hot baryons in Section 3, we then go on in
Section 4 to validate our model predictions for nearby ($z=0$) galaxy
systems against an extensive set of X-ray observations involving a
variety of both scaling relations between bulk properties and
structural properties of the IHM, discussing the results in the light
of recent observations. The last section of the paper contains a
summary of our main findings.

\vspace{1truecm}

\section{THEORETICAL FRAMEWORK}\label{theory}

We present here the set of equations that completely defines our model
for the evolution of the structure of bound X-ray groups and galaxy
clusters. As usual in analytic approaches of this kind, these systems
will be approximated by spherically symmetric structures whose global
dynamics is exclusively driven by the assemblage of the dark matter
component (baryons amount $\lesssim 15\%$ of the total gravitational
mass $M$).

\subsection{The Dark Halo Evolution Model}\label{dark_matter}

Numerical simulations \citep[e.g.,][]{NFW97,Jin00,Bul01} of the
hierarchical mass assembly in the universe, have demonstrated that the
mass density distribution in the $\Lambda$CDM cosmology of relaxed,
nonbaryonic halos \emph{at any redshift} $z$ can be characterized by a
universal NFW functional form
\begin{equation}\label{rhonfw_r}
\rho(r)=\rhoc\frac{\rs^3}{r(\rs +r)^2}\;,
\end{equation}
where $\rhoc$ and $\rs$ are, respectively, the halo characteristic
density and scale radius. 

\citeauthor*{SSM98} \citetext{\citeyear{SSM98}; see also
\citealt*{RGS98}} developed a consistent analytic description of the
hierarchical evolution of DM halos that provides an excellent fit to
$N$-body simulations \citep*{RGS01}. This framework for structure
formation incorporates in the well-known extended Press-Schechter
formalism \citep{LC93} a pre-established phenomenological threshold
$\Dthr$ setting the fractional mass increase that separates the two
basic mass aggregation regimes of dark halos: minor mergers or
accretion, where these objects grow inside-out through the continuous
aggregation of small clumps that do not disturb their internal
structure\footnote{This condition implies that the values of $\rhoc$
and $\rs$ remain unaltered during the inside-out growth process.}, and
major merger events, in which the progenitor halos are fully disrupted
giving rise to the formation of a new bound system. While, as shown by
\citet{RGS01}, it is possible to achieve an excellent agreement
between theory and simulations for any value of $\Dthr$, it is
essential specifying the exact value of this parameter in order to
explain the characteristic inner structure of halos \citep{Man03}, as
well as any correlation involving their concentration \citep{SMS05}. In
particular, the concentration dependence on halo mass at any redshift
found by \citet*{ENS01}, which we shall adopt here, requires
$\Dthr=0.21$ for the elected cosmology \citep{Hio03}. Once $\Dthr$ has
been chosen, the epoch of formation of a population of halos of mass
$M$ ---defined as the redshift at which they have experienced their
last major merger--- is fixed. We take the median value of the
analytical probability distribution of formation times for DM halos
with observed virial masses $M$ \citetext{see eqs.\ [6]--[12] in
\citealt{SSM98}} as their typical formation redshift $\zf$.

By defining $x=r/\rs$ as the radial distance in units of the scale
radius, we can rewrite equation~(\ref{rhonfw_r}) in dimensionless form
as
\begin{equation}\label{rhonfw_x}
\widetilde{\rho}(x)\equiv\frac{\rho(x)}{\rhoc}=\frac{1}{x(1+x)^2}\;,
\end{equation}
where, taking into account the relationship 
\begin{equation}\label{mvir}
M=\frac{4\pi}{3}\rvir^3\Dvir\rhou\;,
\end{equation}
with $\Dvir$ the ratio of the mean density of a sphere of radius
$\rvir$ to the characteristic background density of the universe
$\rhou$ at the epoch in which halos are observed, one has that
$\rhoc=\Dvir\rhou c^3/3[\ln(1+c)-c/(1+c)]$, with $c=\rvir/\rs$ the halo
concentration giving the reduced virial radius of the system within
which the mean halo density is $\Dvir\rhou$. In our model,
$\rhou=\rho_{\mathrm{crit}}(z)=3H^2(z)/8\pi G$ and
$\Dvir(z)=178\Om^{0.45}(z)$, resulting in $\Dvir\approx 100$ at $z=0$.

This dimensionless form of the halo density profile is used to find the
dimensionless mass within radius $x$
\begin{equation}\label{mtilde}
\widetilde{M}(x)\equiv\frac{M(x)}{\Mc}=\ln(1+x)-\frac{x}{1+x}\;,
\end{equation}
the corresponding dimensionless halo circular velocity square
\begin{equation}\label{vtilde}
\widetilde{V}^2(x)\equiv\frac{V^2(x)}{\Vc^2}=\frac{\ln(1+x)}{x}-\frac{1}{1+x}\;,
\end{equation}
and, from the Poisson equation $\rs d\Phi(x)/dx\,$=$\,GM(x)/x^2$ and
expression~(\ref{mtilde}), the dimensionless NFW halo potential
\begin{equation}\label{phitilde}
\widetilde{\Phi}(x)\equiv\frac{\Phi(x)}{\Phic}=-\frac{\ln(1+x)}{x}\;, 
\end{equation}
where the (invariant after halo formation) characteristic parameters
$\Mc=4\pi\rs^3\rhoc$, $\Vc^2=G\Mc/\rs$, and $\Phic=-\Phi(0)=\Vc^2$ act
as normalization constants.

Besides, for the computation of the halo total energy (\S~\ref{ebal})
it will be also necessary to deal with both the reduced one-dimensional
velocity dispersion of the dark matter, which we obtain numerically by
integrating the Jeans equation, assuming isotropic orbits and a null
pressure at infinity,
\begin{equation}\label{Jeans_dm}
\widetilde{\sigma}^2(x)\equiv\frac{\sigma^2(x)}{\Vc^2}=\frac{1}{\widetilde{\rho}(x)}\int_x^\infty \widetilde{\rho}(x)\frac{\widetilde{M}(x)}{x^2}\;dx\,,
\end{equation}
and the masses halos with present-day total masses $M$ had at $\zf$,
which we infer from the rate of halo mass growth by accretion given in
\citet{RGS01}.

\subsection{Modeling the Hot Gas Component}\label{hotgas}

\subsubsection{Dimensionless Profiles}\label{gasprofs}

The equilibrium structure of the IHM is determined by requiring this
component to be, in epochs of gentle mass accretion, in thermal
pressure-supported hydrostatic equilibrium within the (fixed)
gravitational potential wells set by the dark matter halos\footnote{An
implicit and necessary hypothesis of this kind of models is that both
the gas and the dark matter swiftly readjust to a new hydrostatic
equilibrium after a major merger event; see, for instance, the
discussion in \citet*{CMT99}.}. Hydrodynamical simulations on the
physics of diffuse baryons in accretion flows indicate that most of the
bulk energy of the infalling gas is converted through highly efficient
shocks into thermal energy and that the outer boundary of the hot gas
expands following closely the growth of the collisionless dark matter
component \citep[see, e.g.,][]{TM98}. So, as a practical approximation
for the evolution of the IHM structure, we will assume that between
major mergers the hot gas distribution evolves inside-out maintaining
its outermost radius permanently equal to $\rvir$. Furthermore, we will
consider that all the baryonic matter within the virialized halos is in
the hot X-ray-emitting phase \citetext{observational data indicate
that, typically, only $\lesssim 10\%$ of the baryons in clusters are
locked into stars and cold gas so $\Omega_{\mathrm g}\simeq\Omb$; see,
e.g., \citealt*{LMS03,Bel03}}.

The Jeans equation for an ideal gas, $P_{\mathrm X}=\nx
k\tx=\rhox\stx$, with isotropic pressure takes the form
\begin{equation}\label{Jeans}
\stg(x)\left[\frac{d\ln{\rhog(x)}}{dx}+\frac{d\ln\stg(x)}{dx}\right]=-\frac{d\Phi(x)}{dx}\;.
\end{equation}
Thus, to solve the IHM structure by relating univocally the gas
density and temperature to the halo mass more information in the form
of a relation between $\stg$ and $\rhog$ is required. Analytic models
obtain conveniently flexible solutions by adopting for the IHM profiles
a polytropic equation of state $P\propto\rho^\gamma$ (see also
\S~\ref{ebal}), leading to the relationship
\begin{equation}\label{polytrope}
\rhog(x)\propto\stg(x)^{1/(\gamma-1)}\;,
\end{equation}
where $\gamma$ is the polytropic index ---which for a given potential
effectively specifies the shape of the temperature profile--- that is
presumed independent of the halo mass and radial distance to the halo
center. Since we are adopting the common approximation of describing
the hot X-ray plasma by a single-phase (i.e., each volume element
contains gas at just a single temperature), monoatomic perfect gas with
a null metallicity gradient, for a strictly adiabatic IHM $\gamma$ must
be equal to 5/3, the ratio of specific heats. Gas models with a
polytropic index larger than this value are convectively unstable,
while isothermal distributions are retrieved for $\gamma=1$ (apart from
the cores of some clusters, there is little observational evidence for
radially increasing X-ray temperatures, corresponding to $\gamma<1$).

The solution of the equilibrium equation (\ref{Jeans}) by applying
boundary conditions at the current value of the virial radius of the
halo (in $\rs$ units), i.e., at $x=c$, leads to the following
dimensionless temperature profile for $0\le x\le c\;$:
\begin{equation}\label{tg}
\tstg(x)\equiv\frac{\stg(x)}{\Vc^2}=\tstg(c)+\frac{\gamma-1}{\gamma}\left(\widetilde{\Phi}(c)-\widetilde{\Phi}(x)\right)\;.
\end{equation}

Therefore, from equations~(\ref{polytrope}) and (\ref{tg}) the
dimensionless gas density writes
\begin{eqnarray}\label{rhog}
\trhog(x)\equiv\frac{\rhog(x)}{\rhogc}=\trhog(c)\left(\frac{\tstg(x)}{\tstg(c)}\right)^{1/(\gamma-1)}= {~~~~~~~~}\nonumber\\ 
\trhog(c)\exp\left\{\frac{1}{\gamma-1}\ln\left[1+\frac{\gamma-1}{\gamma}\ln\left(\frac{\trhogiso(x)}{\trhogiso(c)}\right)\right]\right\},
\end{eqnarray}
where $\rhogc=\rhoc\Omb/\Om$ is the characteristic density of the gas and 
\begin{equation}\label{rhog_iso}
\trhogiso(x)=\trhogiso(c)\;\exp\left(\frac{\widetilde{\Phi}(c)-\widetilde{\Phi}(x)}{\tstg(c)}\right)
\end{equation}
represents the solution of the Jeans equation for an isothermal gas
with a dimensionless temperature whose value is equal to
$\tstg(c)$. Equations (\ref{tg})--(\ref{rhog_iso}) completely determine
the structure of the hot gas once its temperature and density at $x=c$
are specified.

\subsubsection{Boundary Conditions}\label{boundary}

To determine the value of the additive constant $\tstg(c)$ in equation
(\ref{tg}), we have taken into account the fact that, in the absence of
non-conservative processes, the total specific energies of the gas and
dark matter must remain invariant and equal to each other, as numerical
simulations confirm \citep[][]{Tho01,Mua02} ---we are neglecting
the small gas heating that may arise from any energy transfer between
the dark matter and the hot gas occurring during gravitational collapse
\citep*[see, e.g.,][]{PTC94}. Thus, when the preheating energy of the
gas is non-zero, we have, for any redshift $z\leq\zf$:
\begin{equation}\label{Ebal_1}
{\cal E}_{\mathrm g}={\cal E}_{\mathrm{DM}}+\sDE\;.
\end{equation}
In equation~(\ref{Ebal_1}) ${\cal E}_i=(K+U)_i/M_i$ represents the
ratio between the sum of the total kinetic and potential energies and
the total mass of the $i$ component calculated within the reduced
virial radius $c(z)$, while in $\sDE=\DE/\mubpre$ the non-starred $\DE$ is
chosen to represent from now on the excess energy \emph{per gas
particle} brought by preheating and $\mubpre$ the mean mass of the gas
particles where this energy is injected.

By substituting in equation (\ref{Ebal_1}) the model profiles derived
in the current and former sections, we obtain, after some algebra, the 
implicit relationship 
\begin{eqnarray}\label{Ebal_2}
\frac{\int^{c}_0 \left(3\tstg(x)+\widetilde{\Phi}(x)\right)
\trhog(x)x^2dx-2\sDE/\Vc^2}{\int^{c}_0 \trhog(x)x^2dx}= \nonumber\\
\frac{\int^{c}_0 \left(3\widetilde{\sigma}^2(x)+\widetilde{\Phi}(x)\right)\widetilde{\rho}(x)x^2dx}{\int^{c}_0 \widetilde{\rho}(x)x^2dx} \;,
\end{eqnarray}
which we choose to solve at the observed redshift $z=0$ (see also
\S~\ref{ebal}). In this last equation the value of $\mubpre$ is chosen
by assuming that energy injection takes place after reionization is
complete ($z\lesssim 6$), but before the galaxy systems have formed
(recent estimates by \citealt{Fin03} and \citealt{OB03} suggest that
preheating may have happened at $z\lesssim 3$, a value substantially
larger than the typical formation redshifts $\lesssim 1$ our model
predicts for group and cluster-sized halos). Accordingly, we consider
that the excess energy is injected into a fully ionized plasma with
metallicity equal to the current standard $30\%$ solar abundance and
temperature above $10^5\ {\mathrm K}$ ---so $\mu$ becomes temperature
independent--- for which $\mubpre=0.998\times 10^{-24}\ {\mathrm g}$,
as inferred from the tables of \citet{SD93}. Note that, for the
smallest systems, the nongravitational heating energy $\DE$ is
expected to be the best part of the energy budget, making their mean
gas temperature almost independent of the system mass.

To set the normalization $\trhog(c)$ of the dimensionless gas
density profile (eq.~[\ref{rhog}]), we choose to impose the boundary
condition, also adopted in other models with preheating \citep[see,
e.g.,][]{OB03}, $\rhog(\rvir)/\rho(\rvir)=\Omb/\Om$ implying that
\begin{equation}\label{rhog_nor}
\trhog(c)=\widetilde{\rho}(c)\;.
\end{equation} 
The results of both adiabatic simulations and simulations with cooling
and star formation \citep{Asc03,KNV05} show that the local ratio of gas
to dark matter density within present-epoch halos is close to the
cosmic baryon fraction near the halo boundary. Therefore, by using
equation (\ref{rhog_nor}), we implicitly assume that, in the range of
halo masses where our preheating model is aplicable (see
\S~\ref{lxvstx}), the baryon fraction near the virial radius is not
strongly affected by the non-gravitational heating. Besides, a 'local'
constraint ---as opposed to the more widely used 'global' boundary
condition $M_{\mathrm g}=(\Omb/\Om)M$--- allows for the possibility
that the finite central temperature of the IHM (see, e.g.,
eq.~[\ref{tg}]), which causes the central gas distribution to be more
flattened than the density distribution of the dark matter, can produce
baryon fractions within the virial radius smaller than the cosmic
value. Indeed, given that any energy injection into the gas actually
enhances this tendency by creating a core in its density profile and
decreasing further its central value, equation (\ref{rhog_nor}) grants
that our model can reproduce the reduced total gas masses
characteristic of low-mass halos in preheating scenarios
\citetext{\citealt{Mua02,OB03}; see also the last paragraph of
\S~\ref{msimtx}}.

\section{BEST SOLUTION TO THE EQUILIBRIUM EQUATIONS}\label{best_sol}

Before we investigate the X-ray properties of real galaxy systems with
our model, it is necessary to identify the correct solution to the
equilibrium equations of the hot baryons. In our analytical treatment,
the gas structure is controlled by the values of the polytropic index
$\gamma$ and the preheating energy per particle $\DE$, which we now
proceed to determine. The appropriateness of the resulting best model
will then be exhaustively checked in Section~\ref{results} by
confronting its predictions with an ample variety of X-ray data on both
the global properties and radial structure of the gas.

\subsection{Fixing \boldmath $\gamma$ from the Specific Energy Balance}\label{ebal}

While $\gamma$ appears to be a parameter relatively well constrained
empirically, the consensus on its value is by no means absolute. A
number of observations \citep*[e.g.,][]{Mar98,FAD01,FRB01,San03} and
both analytical models \citep[e.g.,][]{CMT99,TN01} and high-resolution
gasdynamical simulations \citep[e.g.,][]{Asc03,Bor04,Ett04}, agree in
showing that the IHM of galaxy clusters is well represented, outside
any cooling region, by a polytropic equation of state with a universal
index of around $1.2\pm 0.1$ over the full radial range. There are,
however, some recent observations suggesting both that the polytropic
treatment of the gas is only acceptable beyond about 20\% of the
cluster virial radius and that the required value of $\gamma$ is close
to 1.5 \citep{DM02}. As we next show, the physical consistency of the
inside-out growth of the halo structure in regard to the energy balance
between its two main components can be used to infer a well defined
value for this index, thereby implying that $\gamma$ is not a true
degree of freedom in our model.

As stated in Section~\ref{boundary}, relation~(\ref{Ebal_1}), which was
used to set the boundary condition on the equilibrium dimensionless
temperature profile, must hold from $\zf$ provided a negligible
fraction of the hot gas is able to cool after the halo formation. While
we have chosen to fix the balance of total specific energies between
the two main halo constituents, ${\cal E}_{\mathrm g}-{\cal
E}_{\mathrm{DM}}$, at the observed redshift $z=0$, obviously there is
no guarantee whatsoever that its universal value, $\sDE$, is preserved
automatically for any other cosmic time larger than the virialization
epoch. In fact, as the equilibrium equations of the inside-out evolving
DM and polytropic gas profiles are different, the degree to which
condition~(\ref{Ebal_1}) is fulfilled along the halo lifetime depends
on the specific values adopted for the parameters $\gamma$ and $\DE$
controlling the gas structure.

\begin{figure}
\centerline{\includegraphics*[width=9.5cm]{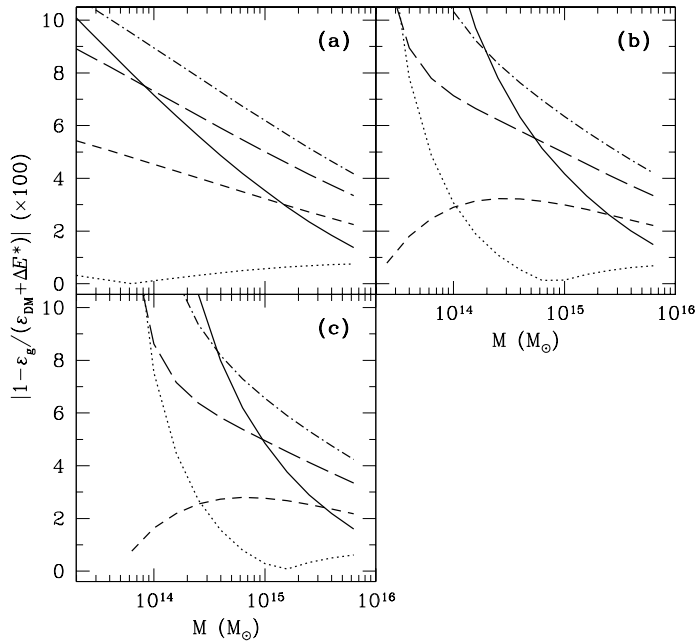}}
\figcaption{Model percent deviation at $\zf$ between the sum of the
kinetic and potential specific energies of the hot gas and its
corresponding total specific energy inferred from
equation~(\ref{Ebal_1}) plotted against the total virialized mass of
present-day halos. Comparison is made for three different levels of
energy injection: $\DE=0$\keVpar\ (panel a), $\Delta E=0.5$\keVpar\
(panel b), and $\DE=1.0$\keVpar\ (panel c). In all panels, different
line types denote different values of the polytropic index: $\gamma=1$
(solid), $\gamma=1.1$ (dotted), $\gamma=1.2$ (short-dashed),
$\gamma=1.3$ (long-dashed), and $\gamma=1.4$ (dot-short-dashed).
\label{engydif}}
\end{figure}

Figure~\ref{engydif} summarizes the (maximum) percent difference
between the sum of the kinetic and potential specific energies of the
hot gas and the total specific energy of this component inferred from
equation~(\ref{Ebal_1}) calculated at $\zf$. Results for three
different levels of mass-independent $\DE$, each one for five different
values of $\gamma$, are presented. We note that the physical
consistency of our predictions has been also preserved by limiting the
calculations to halo masses for which the sign of the sum of the
kinetic and potential energies of the IHM is negative, as otherwise it
cannot remain gravitationally bound (see also
\S~\ref{lxvstx}). Figure~\ref{engydif} shows that, in general, the
deviations increase with decreasing present-day gravitational mass
reflecting the more important contribution of the extra heating to the
gas energy budget in small halos. In panel (a), where the results
corresponding to a null preheating ($\DE=0$) are displayed, the value
of $\gamma=1.1$ gives the best performance, keeping the relative
deviation below 1 per cent across the entire range of halo masses that
are being investigated. The curve for $\gamma=1.2$ also shows
moderately low deviations ranging from 2\% at the high mass-end up to
somewhat less than 6\% for the smallest systems. Compared to these two
models, the performances of the $\gamma=1.3$ and $\gamma=1.4$ solutions
are much less satisfactory. This latter conclusion is also applicable
to the results obtained for an isothermal gas profile ($\gamma=1$) in
the low half of the mass range, which only leads to results as good as
in the $\gamma=1.1$--1.2 cases for halo masses above $\sm
10^{15}$\msun.

The effects of including nongravitational gas heating are illustrated
in panel (b) of Figure~\ref{engydif}, were we display the deviations
calculated for $\Delta E=0.5$\keVpar, and panel (c), were a larger
amount of energy injection of $1.0 $\keVpar\ has been assumed. The
addition of preheating energy results in larger discrepancies, except
for the case $\gamma=1.2$, which now preserves the universality in the
balance of specific energies to within 3\% across the
three-order-of-magnitude range of gravitational masses. Now, even the
behavior of the $\gamma=1.1$ curves, which is highly satisfactory for
cluster-scale systems, swiftly deteriorates once they move below $\sm
1$--$2\times 10^{14}$\msun, giving rise to unacceptably high energy
differences. Under these same conditions, the relative deviation of the
isothermal solutions remains satisfactorily low only for the rarest
most massive clusters. We note that all curves in Figure~\ref{engydif}
have been calculated by taking $\zf$ equal to the median of the
formation time distribution of the dark halos of a given current mass
$M$. In general, the larger the adopted formation redshift, the larger
the maximum relative deviation that results at this redshift. Thus, for
instance, by setting $\zf$ equal to the lower quartile the
discrepancies become typically a factor two larger.

These results all together imply that for a moderately preheated IHM
$\gamma \sm 1.2$ warrants the best global physical consistency with the
inside-out growth of structure by accretion. Interestingly, polytropic
gas distributions with $\gamma=1.1$--1.2 in a NFW potential well also
have the desirable property of tracing the dark matter profile in the
outer part of halos \citep{KS01}. Besides, as we discuss in the next
section, a universal polytropic index of 1.2 for the diffuse baryons
provides the best fit of our predictions to the observations when X-ray
data are used to adjust $\gamma$ and $\DE$ \emph{simultaneously}.

\subsection{Constraining $\DE$ from the Luminosity-Temperature 
Relationship}\label{lxvstx}

The X-ray luminosity and temperature are the most easily observable
bulk properties of the hot gas in clusters and groups of
galaxies. Accordingly, they have been measured for a large number of
these systems. This large dynamic range of the data and the few
hypothesis involved in the calculation of these two observables, which
are reflected in a relatively tight correlation, render this
relationship the most reliable empirical reference for the calibration
of any theoretical modeling of the IHM. In this section, we fix the
poorly constrained value of $\DE$ by finding the theoretical solution
that, for $\gamma=1.2$, best matches the observed
luminosity-temperature relationship for groups and clusters of
galaxies.

Comparison to observations is done via the total \emph{bolometric}
X-ray luminosity, derived by integrating the volume emissivity of the
X-ray-emitting gas, $\epsilon(x)$, out to the halo maximum radius:
\vspace{0.5truecm}
\begin{eqnarray}\label{lx}
\lx&=&4\pi\rs^3\int^c_0 \epsilon(x)x^2dx \nonumber \\
     &=&4\pi\rs^3\rhogc^2\int^c_0 \Lambda(\tstg,Z)\left(\frac{\trhog(x)}
{\mub(\tstg,Z)}\right)^2x^2dx\;,
\end{eqnarray}
and the emission-averaged X-ray temperature of the halo \emph{expressed
in energy units}
\begin{equation}\label{tx}
\tx=\frac{\Vc^2\int^c_0 \mub(\tstg,Z)\tstg(x)\epsilon(x)x^2dx}{\int^c_0 \epsilon(x)x^2dx}\;.
\end{equation}
In equations (\ref{lx}) and (\ref{tx}), $\trhog(x)$ and $\tstg(x)$ are,
respectively, the mass density and temperature profiles of the IHM
given by equations (\ref{tg})--(\ref{rhog_iso}), whereas $\Lambda(T,Z)$
is the bolometric cooling function normalized to the total number
density of all plasma species (electrons and ions) taken from the
tables of \citet{SD93}, which include both bremsstrahlung and cooling
from metal lines, for a plasma of metallicity $Z=0.3\;Z_\odot$. In
these calculations, we have kept track of the cooling time at $c$,
\begin{equation}
t_{\mathrm{cool}}=\frac{3}{2}\rhogc\Vc^2\frac{\trhog(c)\tstg(c)}{\epsilon(c)}\;,
\end{equation}
in order to exclude halos for which all of the gas within the
virial radius has had time to cool since the system formed. 

\begin{figure}
\centerline{\includegraphics*[width=9.5cm]{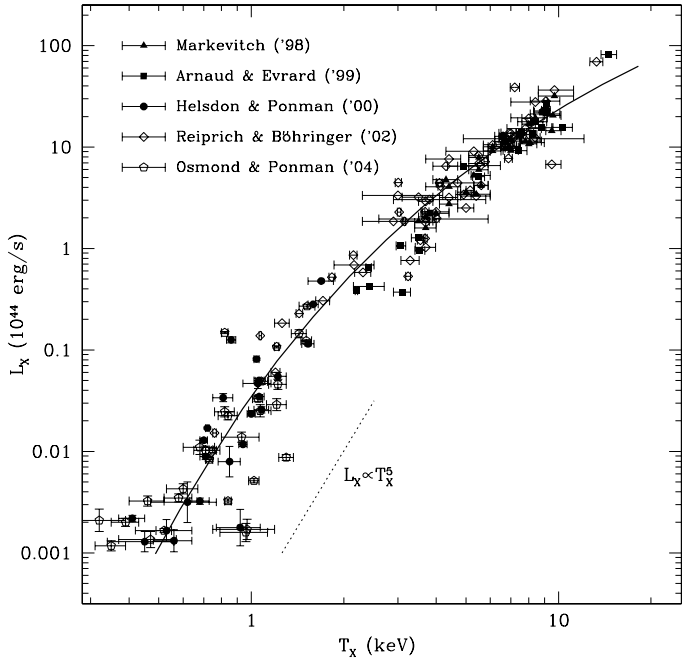}}
\figcaption{Luminosity-temperature relation versus
observations. Different datasets are identified with different
symbols. The solid line shows our best-fit model prediction drawn for a
polytropic gas with $\gamma=1.2$ and metallicity equal to
$0.3\;Z_\odot$, which yields $\DE=0.55$\keVpar. A Hubble parameter
of $h=2/3$ has been applied to both the model and the data.
\label{lxtx}}
\end{figure}

Figure~\ref{lxtx} presents the comparison of the observed $\lx-\tx$
relation for groups and clusters of galaxies with our best-fit model,
which is obtained for $\DE=0.55$\keVpar. In this figure, the
theoretical curve encompasses halos of present-day virial masses
ranging from $6\times 10^{15}$\msun\, large enough to explain the
hottest system data, down to $2\times 10^{13}$\msun. The exclusion of
systems with masses below the latter value is due to the impossibility
for halos this small of retaining their IHM when energy injection
reaches 0.55\keVpar. Interestingly the emission-weighted X-ray
luminosity associated with this lower-mass limit ($\sm 10^{41}$\ergss)
shows a good correspondence with the minimum luminosity of the
measurements in the \emph{GEMS} galaxy group project by \citet{OP04}
when this dataset is restricted to those systems with genuine group
emission (G-sample; open pentagons in the figure).

Figure~\ref{lxtx} also illustrates that the slope of our best-fit model
increases gradually with decreasing $\tx$ matching the trend of the
observations at cluster and group scales. In the hot cluster regime
($\tx\gtrsim 3$ keV) $\lx\propto\tx^{\sim 3}$, as suggested by most
authors \citep*[e.g.,][]{WJF97,AE99}, although our model indicates that
the logarithmic slope should approach asymptotically the self-similar
value of 2 for the very hottest systems. For the coolest systems, the
typical model behavior ($\lx\propto\tx^{\sim 5}$) is likewise fully
consistent with observations, including the latest data from the
\emph{GEMS} project G-sample that, nonetheless, show a noticeably
increase in the scatter about the predicted trend for $\lx\lesssim
10^{42}$\ergss. As argued by \citeauthor{OP04}, this marked raising of
the dispersion ---arising, for the most part, from pushing the current
instrumental capabilities to the limit in order to observe a
sufficiently representative number of cool groups ($\tx < 0.7$ keV)---
introduces several sources of bias that likely conspire to flatten
the $\lx-\tx$ relationship at these scales and render it to apparently
follow the cluster trend, instead of the characteristic logarithmic
slope of 4--5 found for X-ray bright groups \citep{HP00,XW00}. Still,
the fact that our model goes through the middle of the locus occupied
by the data points in the low $\tx$, low $\lx$ regime suggests that the
$\lx-\tx$ relationship does indeed steepen substantially in these
extreme scales.

It is also interesting to note that had we decided not to fix for a
start the value of $\gamma$ by imposing the preservation of the
specific energy balance between the two major halo components during
the accretion phase, but use instead the $\lx-\tx$ relationship to
adjust simultaneously the values of this parameter and of the excess
energy, we would have inferred, precisely, a best-fit value of 1.2 for
the universal polytropic index. This argues in favor of the consistency
of the X-ray data with the inside-out growth of the structure of galaxy
systems within a universal preheating scenario in which $\DE$ is about
half a keV per particle.

\begin{figure*}
\centerline{\includegraphics*[width=13.5cm]{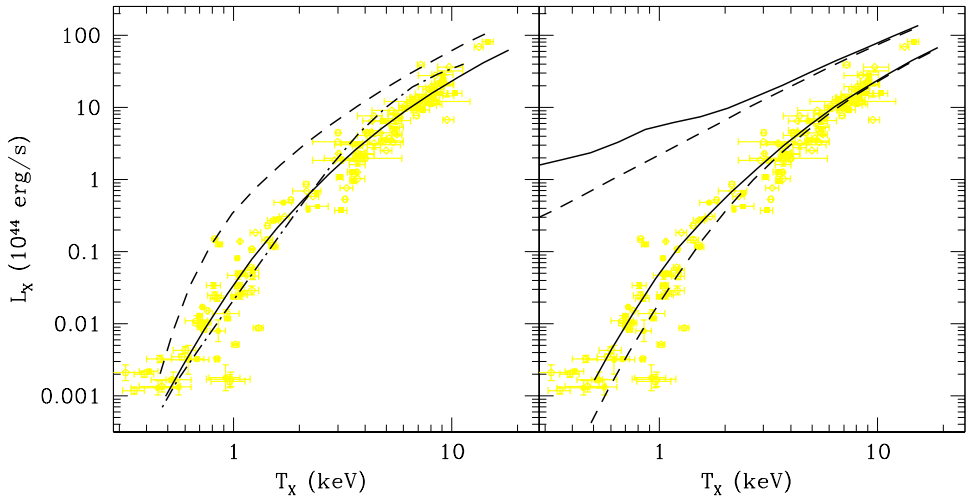}} \vspace{-6cm}
\figcaption{\emph{Left:} Comparison between our best solution (solid
curve) with the predictions for an isothermal ($\gamma=1$;
short-dashed curve) and nearly adiabatic ($\gamma=1.5$;
dot-short-dashed curve) gas endowed with the same amount of injected
energy. \emph{Right:} Effects of metallicity on the
luminosity-temperature relation. The two lowest curves are from a
$\gamma=1.2$, $\DE=0.55$\keVpar\ model with zero (short-dashed
line) and solar (solid line) gas metallicities. The two upper curves
are the corresponding predictions resulting from a non-radiative
$(\gamma,\DE)=(1,0)$ model (same coding as before). The meaning of
symbol types is the same as in Fig.~\ref{lxtx}. Note how the curves
from the model with null energy injection clearly fail to match the
observational data.
\label{lxtxparams}}
\end{figure*}

On the other hand, the fact that our model shows the closest agreement
to the observations across several orders of magnitude in X-ray
luminosity for a value of the preheating energy that compares favorably
to constraints found in the literature is very assuring and can be
viewed as a further endorsement of the consistency of our
approach. Thus, our best-fit value of $0.55$\keVpar\ for this parameter
agrees very well with the energy budgets inferred from recent
analytical and numerical IHM models \citep{Voi02,Fin03}, as well as
with the results of attempts at directly measuring the extra energy
injected into the gas \citep{LPC00}. To our knowledge ours is among the
universal preheating models with the lowest heat input requirements. In
this respect, it is worth noting that excess energies of only half a
keV per particle might not be out of the reach of the total energy
released into (though not necessarily retained by) the intergalactic
medium by galaxy winds in starburst galaxies or by active galactic
nuclei (AGN), to name only a few of the most popular heating sources,
which comes to several keV per particle when averaged over all baryons
in the universe \citep*[e.g.,][]{CLM02}.

\subsection{Sensitivity to Model Parameters}

While specifying $\DE$ completely determines the gas profiles, it is
reasonable to wonder how our predictions depend on the exact values
adopted for this parameter, for the fixed-by-theoretical-arguments
polytropic index, as well as for the other parameters of the model that
have been prefixed for definiteness. In the following we comment on
this issue, focusing on the $\lx-\tx$ relationship.

As was already noted on Section~\ref{boundary}, the effects of the
energy injected into the gas at preheating become progressively
noticeable with decreasing halo mass (temperature), reflecting the
mounting contribution of $\DE$ to the total energy of this
component. In particular, we find that the behavior of the predicted
luminosity-temperature relationship below $\tx\sim2$ keV becomes
extremely sensitive to the exact value adopted for this degree of
freedom, to the point that in the adjustment process the X-ray data
allows one to easily discriminate between variations of $\DE$ larger
than only $0.05$\keVpar\ in spite of the fact that, at these scales,
the dispersion of the observations is substantial.

While $\Delta E$ sets the normalization and convexity of the
luminosity-temperature relationship at the low-mass end, the amplitude
of this correlation at cluster scales depends essentially on the value
of the polytropic index. By varying $\gamma$ (at the cost of lessening
the physical consistency of the predictions), we have found that
relatively small changes of one tenth around its preferred value of 1.2
produce theoretical $\lx-\tx$ curves that differ only marginally from
the best solution. The results of Section~\ref{ebal} demonstrate,
however, that this mild degeneracy vanishes completely when the
preservation of the balance between the total specific energies of the
gas and dark-matter components after halo formation for the entire
range of scales we are considering is taken into account. In contrast,
as the left panel of Figure~\ref{lxtxparams} illustrates, when the
extremal values of the polytropic index $\gamma=1$ (isothermal IHM) and
$\gamma=1.5$ (nearly adiabatic IHM) are combined with our best-fit
value for the injected energy, the model predictions for the adopted
$\Lambda$CDM cosmogony clearly fail to match the X-ray data. It is
worth noticing, however, that our solutions for $\tx\lesssim 2$ keV are
relatively insensitive to the value of $\gamma$, as far as it is larger
than $\sm 1.15$. This would explain why injection models adopting
$\gamma=5/3$ for the gas are successful in describing the
low-luminosity group data in spite of the lack of observational support
for an adiabatic gas distribution at these scales. Moreover, the fact
that the isothermal solution runs essentially parallel to our best
model across most of the cluster regime, implies that universal
preheating models relying on the isothermality of the IHM can be
matched up with the data if they are endowed with some additional
freedom that allows one to renormalize, directly or indirectly, the
X-ray luminosity (temperature).

Other factors that can influence our model predictions are those
related to the formation history of DM halos, such as the underlying
cosmogony, specified by the values assumed for the cosmological
parameters, and the adopted mass-concentration relation ---we have
verified that the predicted gas properties for our best model remain
essentially independent of the adopted halo formation epoch even when
an explicit dependence on it is introduced by solving equation
(\ref{Ebal_2}) at $\zf$. With regard to the dark and baryonic matter
contributions to the density parameter, any variation in $\Omb$ and/or
$\Om$ that lowers the cosmic baryon ratio would lower the amount of
baryons in the IHM and thereby the halo luminosities (and, of course,
the hot gas mass fraction; see \S~\ref{msimtx}). At the same time, the
normalization of the perturbation spectrum, $\sigma_8$, has
repercussions on the convexity of the $\lx-\tx$ relationship at cluster
scales, which increases when the value of this parameter is lowered and
vice versa. We also want to emphasize the fact ---already stressed by
\citet{BBP99}--- that, in models relying on the hydrostatic equilibrium
of gas in the dark matter potential, the ratio of model to observed
luminosity is $h$ dependent, so the best-fit solution varies with the
assumed value of the Hubble constant. We have nonetheless checked that
if we take $h=1/2$ then the preferred value of $\DE$ varies by less
than $10\%$, provided we allow for variations on the other cosmological
parameters within the \emph{WMAP} uncertainties. Last but not least, we
also have verified that the adoption in our calculations of the
mass-concentration prescription by \citet{Bul01} produces essentially
identical results.

The non-statistical noise of the $\lx-\tx$ and any other scaling
relationships involving bulk properties of the IHM may be increased by
the intrinsic scatter of the mass-concentration relationship
\citetext{e.g., \citealt{AC02}; see also Fig.~1 in \citealt{WFN00}}, as
well as by deviations of real systems from the ideal conditions of the
modeling (e.g., non-negligible ellipticity, incomplete dynamical
relaxation, or variations in the efficiency of energy injection). While
these are effects that are not considered in the present study, the gas
metallicity is a factor that may introduce a substantial amount of
scatter in the $\lx-\tx$ relationship, especially at group scales,
which is very straightforward to investigate within the context of our
model. It is well established
\citep[e.g.,][]{Ren97,AF98,Fuk98,EAF01,Toz03} that rich clusters
exhibit almost constant average metallicity of $\sm 0.3\;Z_\odot$ with
a reduced scatter, while cooler systems with $T\lesssim 1$ keV show a
wide range of values going from negligible metal contents to solar
abundances ---we do not consider here the existence of abundance
gradients in cluster and groups of galaxies; see, e.g., \citet{DeG04}
and references therein. The fact that metal lines begin to provide a
significant contribution to the emissivity only at temperatures below a
few keV has been frequently exploited in investigations of the X-ray
properties of halos by computing the cooling function under the simpler
conditions that describe the physics of high temperature plasmas
\citep[but see][]{DKW02}. However, when one seeks to extend the
predictions to the group scales, the effects of metallicity can no
longer be overlooked. As we show in Figure~\ref{lxtxparams}
(\emph{right}), the contribution of metal lines to cooling starts to be
noticeable, just like it happens with the contribution of $\DE$ to the
gas energy, when the IHM temperature drops under 2--3 keV, to the point
that, independently of the amount of non-gravitational heating, the
bolometric X-ray luminosities at $\tx\sm 0.5$ keV inferred from the
model solutions that use a solar-metallicity cooling function become
about three times higher than those in which primeval abundances have
been adopted. (The comparison between the two sets of preheated and
non-preheated solutions drawn in the figure also illustrates the strong
downward bending in the luminosity-temperature relationship that
results from an energy injection of $0.55$\keVpar, which for 0.5 keV
halos implies a reduction of more than two orders of magnitude in $\lx$
for $\gamma=1.2$.) Also note how within the isothermal models
($\gamma=1$) with no energy injection ($\DE=0$), the upper curve, which
represents the solution for a gas with a solar metal abundance, veers
away from the roughly self-similar scaling of the zero-metallicity
solution, showing a shallower, rippled slope that reflects the rises
and falls of the cooling function arising from different metal lines
\citep[see, e.g., Figures 8 and 18 in][]{SD93}.

Lastly, we would like to comment on the fact that some analytical
models also resort to surface brightness bias, i.e., to measure
temperatures and luminosities above an arbitrarily fixed surface
brightness level, to explain both the offset toward lower luminosities
and the increase of the dispersion of the luminosity-temperature
relation for the less massive halos \citep[e.g,][]{Voi02}. Given the
outward-decreasing equilibrium profiles of the gas, any reduction in
the outer halo boundary has the effect of lowering $\lx$ and raising
$\tx$, producing the same trend seen in the data. Although this effect
has not been accounted for in our modeling to avoid introducing an
additional, poorly constrained, degree of freedom, we point out that
raising the surface brightness threshold would have reduced further the
amount of preheating energy necessary to reproduce the observations.

Given that the (relatively important) impact of possible variations in
metallicity is noticeable for a range of system temperatures in which
X-ray data ---other than the X-ray luminosities and temperatures--- are
scarce and affected by large uncertainties, the discussion of the next
section focuses only on the model predictions inferred for halos whose
IHM has a metallicity fixed to the conventional value of
$0.3\;Z_\odot$.

\section{RESULTS}\label{results}

\begin{figure}
\centerline{\includegraphics*[width=9.5cm]{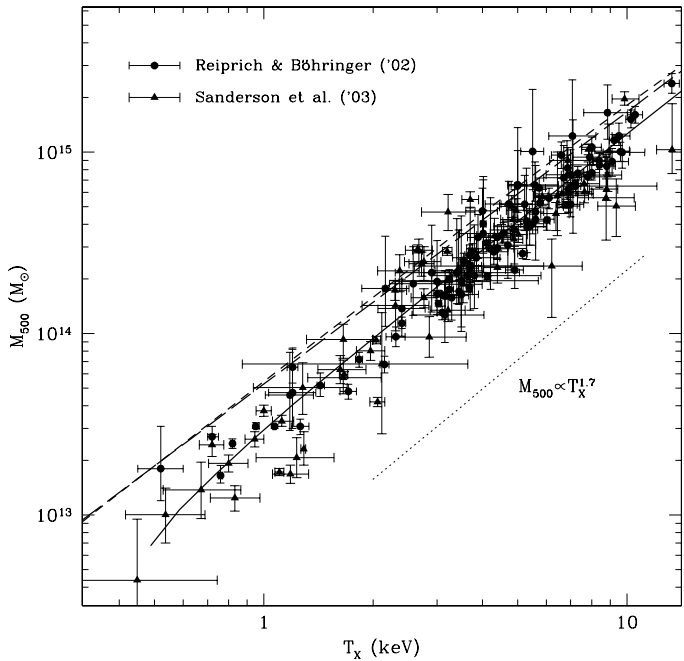}}
\figcaption{Mass-temperature relation versus observations. Different
symbols are used to identify different datasets. The lower solid line
shows the predictions corresponding to the best-fit model in
Fig.~\ref{lxtx}. The two upper curves illustrate that the correlation
between halo mass and gas temperature inferred in the $\DE=0$
simulations of \citet{EMN96} for a IHM obeying an isothermal
($\gamma=1$) $\beta$-model with an iron abundance of 0.3 solar 
(long-dashed line) is closely matched by our corresponding model
prediction (short-dashed line).
\label{m500tx}}
\end{figure}

\subsection{The mass-temperature relationship}\label{msimtx}

Plotted in Figure~\ref{m500tx} is a comparison between the
mass-temperature relationship resulting from our preferred model and
data from both the enlarged \emph{HIFLUGCS} sample \citep{RB02} and the
Birmingham-CfA cluster scaling project \citep{San03}. The temperature
axis shows, as before, the emission-weighted temperature within the
virial radius, $\tx$, while to facilitate comparison with observations,
the halo mass is now represented by $\msim$, the total mass inside a
scaled radius of $\csim$ within which the mean matter density is
$500\rho_{\mathrm{crit}}$. As \citet{San03} actually quote
emission-weighted temperatures within a scaled radius of $0.3\ctwohun$
their values have been conveniently rescaled by means of the
relationship $\tx=\tx(0.3\ctwohun)/1.25$, which we have calibrated from
the comparison between their estimates of the emission-weighted
temperatures for the subsample of \citet{LPC00}'s systems and the
measurements published in this previous study.

The visual inspection of Figure~\ref{m500tx} shows that our predicted
mass-temperature relationship has the right normalization and that the
constant logarithmic slope of $\sm 1.7$ of the model solution above
$\sm 1$ keV is fully compatible with the trend shown by the
observations. This good match between the observed and theoretical
$\msim-\tx$ relations is very compelling, especially considering that
there is no freedom in our prediction. Our results are also in
excellent agreement with previous findings of a logarithmic slope of
$1.79\pm 0.14$ by \citet{NMF00}, by using overdensities of 1000 and
spectroscopically derived temperatures, and of $1.78\pm 0.09$ by
\citet{FRB01}, derived from spatially resolved X-ray observations. In
contrast, the slope of $1.84\pm 0.06$ found by \citet{San03} for their
sample (using $M_\mathrm{200}$) differs by $2.3\sigma$ from our
value. However, a major difference of this latter sample, compared to
those used in the studies previously mentioned, is the inclusion of a
substantial number of systems with temperatures below 1 keV, including
two measurements on early-type galaxies (the lower-leftmost data point
in Fig.~\ref{m500tx} represents the most massive of them, NGC 6482),
which tend to increase the slope of a global log-log linear fit. On the
other hand, the inferred uniform rising of the mass over most of the
temperature range does not support claims of a mass-temperature
relationship that is convex \citetext{\citealt{San03}; but see our
comments below on the total gas mass-temperature relationship} or has a
break at about 3--4 keV \citetext{\citealt*{NMF00};
\citealt{FRB01}}. At this point, we would like to stress that any
conclusions on the actual value of the slope of the mass-temperature
relationship and its possible convexity should be regarded with
caution, mostly because the results being debated are not only
sometimes based on ill-defined, incomplete datasets, but also due to
the different prescriptions, extrapolations, and other inherent
difficulties involved in the calculation of integrated halo properties,
particularly the virial mass. The definitive answer will have to await
the forthcoming availability of larger samples of high-quality data
with well-defined selection procedures which can be accounted for
during the fitting process. What is clear from our analysis is that the
predictions of our universal preheating model compare very favorably to
current observations.

For low-temperature systems ($\tx\lesssim 1$ keV) the predicted
$\msim-\tx$ relationship initiates a break away from the general
$\msim\propto\tx^{\sim 1.7}$ trend, barely perceptible because of the
cut off in mass (and temperature) imposed by the condition that the IHM
must be bound for excess energies of 0.55\keVpar. Had we not taken into
account this restriction the theoretical mass-temperature relation
would had then shown a progressively marked decline, leading to an
emission-weighted temperature of the gas almost independent on the halo
mass, much as the universal preheating model by \citet{BBP99} and
\citet{Bab02} predicts it should happen for a range of masses $\msim$
around $10^{13}$\msun. Most interestingly, this radical variation of
the correlation trend would show a very good correspondence in the
diagram with the locus occupied by the measurements of the two
early-type galaxies included in the Birmingham-CfA cluster sample. This
coincidence is, nonetheless, surprising since, for the best-fit value
of the preheating energy, our model gives rise to galaxy-sized halos
that cannot retain their hot gas.

Concerning the pure preheating model by Balogh, Babul, and
collaborators, we want to stress the fundamental difference existing in
the explanation put forward for the origin of the observed dispersion
in the scaling relationships, which these authors attribute primarily
to the distribution of halo formation times. This conclusion appears to
be a consequence of their assumption that the accretion of the gas on
the low-mass halos is specified by the adiabatic Bondi accretion rate,
which leads to a strong dependence of the gas density normalization on
$\zf$ that propagates, throughout the Jeans equation, to the bulk
properties of this component. In contrast, the properties of the X-ray
gas in our best model are independent of the distribution of halo
formation redshifts. Therefore, taking into account both that, on
cluster scales, the other intrinsic dispersion factors that might
affect the group data are not an issue and that the level of scatter of
the measurements depicted in Figure~\ref{m500tx} is pretty uniform
across the range of observed masses, we attribute the observed scatter
in the $\msim-\tx$ relationship a genuinely statistical nature.

We have also included in Figure~\ref{m500tx} the self-similar relation
derived from the hydrodynamical simulations by \citet{EMN96}, rescaled
to our value of $h$ (long-dashed curve). Both the higher mass
normalization and the canonical 1.5 slope are faithfully described by
our model prediction corresponding to an isothermal gas without
preheating (short-dashed curve). This result, apart from reinforcing
confidence in our model, demonstrates that essentially all the
steepening, as well as the downward shift in the normalization, are 
attributable to the effects of preheating.

\begin{figure}
\centerline{\includegraphics*[width=9.5cm]{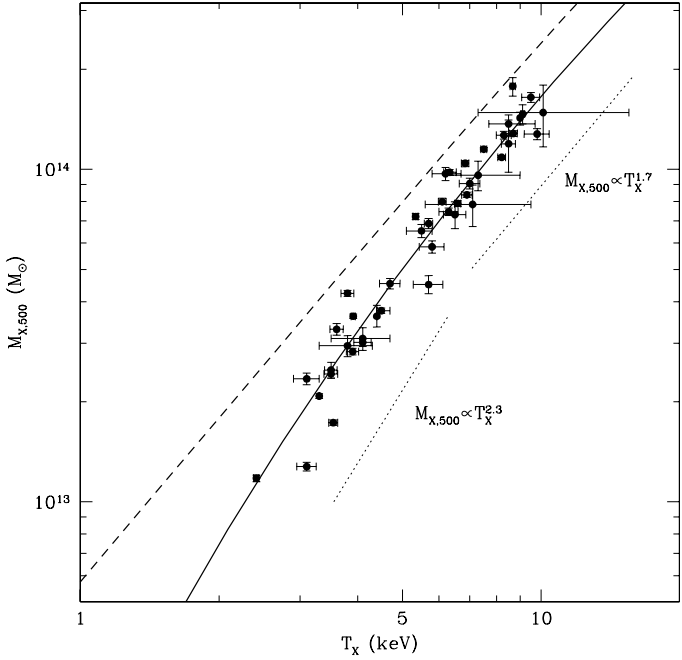}}
\figcaption{Gas mass-temperature
relation versus observations. Solid circles and error bars represent
gas mass measurements within a scaled radius of $\csim$ by
\citet{MME99}. The lower solid line shows the prediction corresponding
to the best-fit model in Fig.~\ref{lxtx}. The upper short-dashed
line shows the prediction resulting from a model with $\gamma=1$ and
$\DE=0$.
\label{mx500tx}}
\end{figure}

The impact of preheating on total gas mass observations can be gauged
in Figure~\ref{mx500tx}, where we depict emission-weighted temperature
versus gas mass $\mxsim$ inside the scaled radius $\csim$. Again, the
trend exhibited by the data, now coming from \citet*{MME99}, is
dutifully reproduced by our relationship. Another remarkable result is
that we predict a convex correlation in which the slope smoothly
increases with decreasing temperature as in the purely analytical model
by \citet{DD02}. The comparison of this correlation with the result for
an isothermal model in which preheating has been switched off
(short-dashed curve), shows that both predictions follow parallel
trends for systems with $\tx\gtrsim 5$ keV (the normalization of the
isothermal model always being higher). For halos this hot, the inferred
typical logarithmic slope is $\sm 1.7$, close to the self-similar value
but nevertheless steeper, and an almost perfect match to the best-fit
slope of $1.71\pm 0.13$ found by \citet{VFJ99} for regular galaxy
clusters. Below this temperature, as the role of preheating becomes
progressively important, the gas mass-temperature relationship of our
best model tends to steepen promptly, adopting power-law indexes
between 2.3--2.5. Not surprisingly, the best-fit linear log-log
correlation determined by \citet{MME99} to their data produced a slope
of $1.98\pm 0.18$, intermediate between these two behaviors.

The fact that our $(\gamma,\DE)=(1,0)$ solution exhibits a nearly
constant slope (there is only a hint of convexity) somewhat steeper
than the self-similar value of 1.5 does not come as a surprise if one
considers that the behavior of the $\mxsim-\tx$ relationship reflects
those of the ratios $\tx/\tvir$ and $\mx/M$. While the
identity between the effective gas and dark matter virial temperatures
and the proportionality of the density distributions of these two main
components for all radii are common assumptions in the modeling of
X-ray halos, our results illustrate that they are only convenient
approximations. Thus, within our scheme, even an isothermal gas with
null non-gravitational heating (otherwise the gas temperature is pushed
well above $\tvir$), leads to values of the ratio between the X-ray
luminosity-weighted and virial temperatures that, in general, border
on, but are not equal to the unity (indeed, in all the solutions
investigated this ratio shows a monotonic increase with increasing halo
concentration). On the other hand, the hotter systems tend to be more
gas-rich than the cooler ones, a tendency that preheating accentuates
by preventing gas from reaching a high density in the central regions
\citetext{\citealt{BEM01,MBB02}; see also Sec.~\ref{profiles_3D}}.

\begin{figure}
\centerline{\includegraphics*[width=9.5cm]{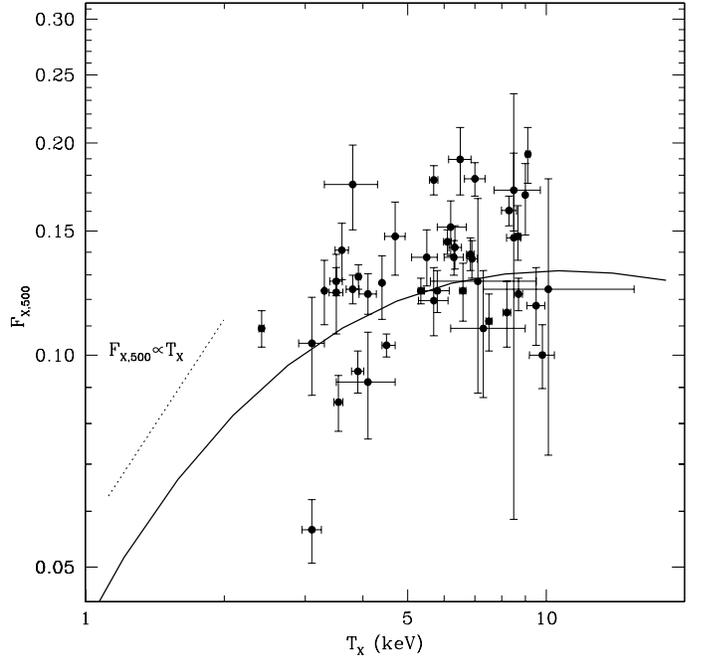}}
\figcaption{Relation between the
integrated gas mass fraction and gas temperature. The data points are
gas mass fraction measurements within a scaled radius of $\csim$ from
\citet{MME99}. The solid curve is the prediction
corresponding to the best-fit model in Fig.~\ref{lxtx}.
\label{fx500tx}}
\end{figure}

\begin{figure*}
\centerline{\includegraphics*[width=14.5cm]{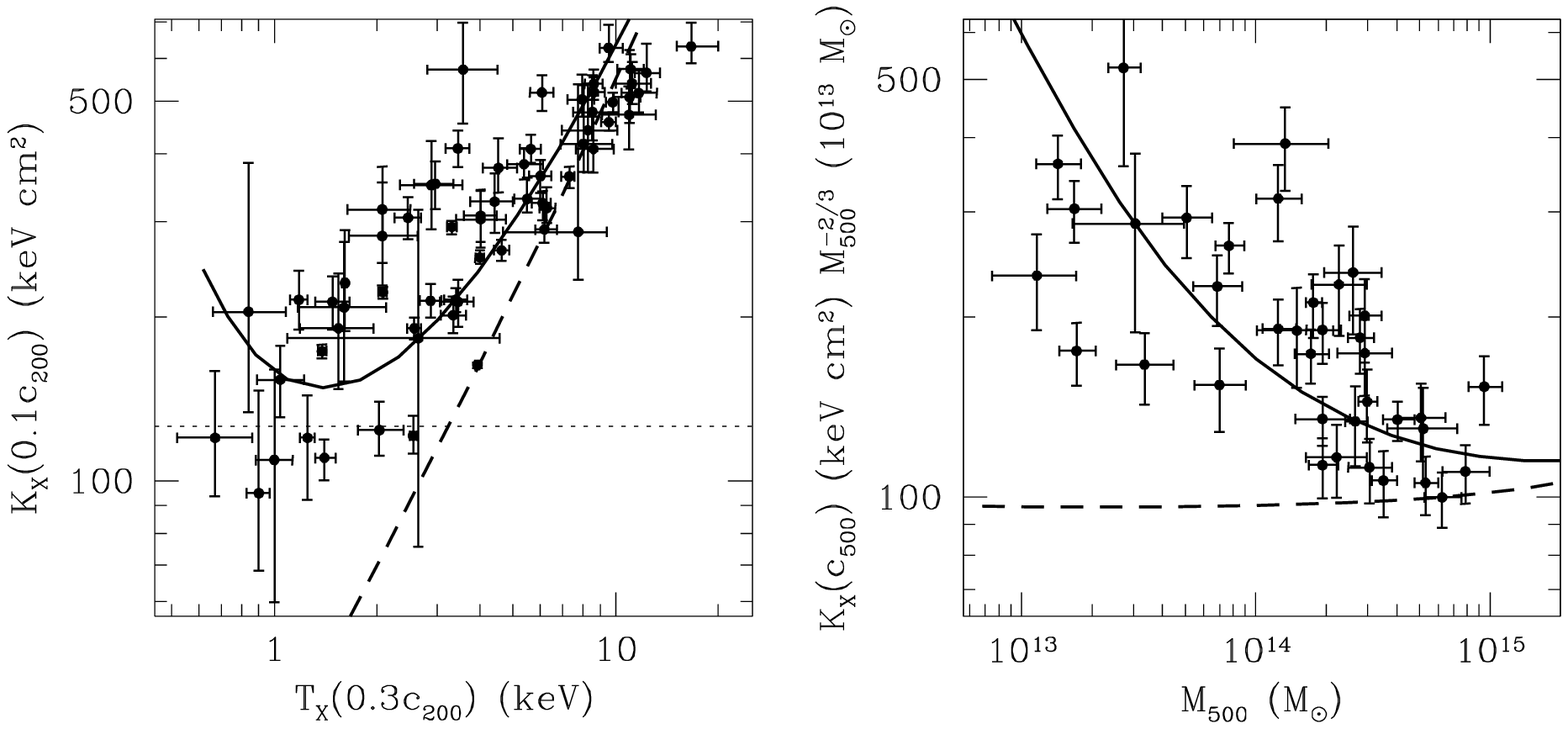}}
\vspace{-7cm}
\figcaption{\emph{Left:} Inner halo entropy at one tenth of
$r_\mathrm{200}$ plotted against the system mean temperature within
$0.3r_\mathrm{200}$. Data points are measurements from
\citet{PSF03}. The right-most solution (short-dashed line) indicates
the effect of switching off preheating (i.e., taking $\DE=0$). The
entropy floor of 126\keVcm\ from \citet{LPC00} is indicated by a thin
horizontal dotted line. \emph{Right:} Scaled outer halo entropy at the
overdensity of 500 vs.\ total mass measured within the same
overdensity. Data points show measurements compiled by
\citet{Fin02}. The near horizontal short-dashed curve shows the solution
when $\DE=0$. In both panels the curves with the largest amplitude
(solid lines) are the predictions corresponding to the best-fit
model in Fig.~\ref{lxtx}.
\label{adiabat}}
\end{figure*}

Finally, we have compared in Figure~\ref{fx500tx} the temperature
dependence of the gas mass fraction as predicted by our preferred model
---calculated by simply dividing the previously inferred masses
$\mxsim$ and $\msim$ as a function of the emission-weighted
temperature--- with data from \citet{MME99}. Looking at this plot one
sees that, regardless of the considerable amount of scatter in the
measurements, the prediction corresponding to our preferred model again
explains the observations remarkably well. One can also observe that
our model prognosticates a progressive steepening of the slope, from
cluster to group scales, which we cannot validate because of the
insufficient dynamical range of the data available \citetext{but note
the apparent consistency of our prediction with the trend delineated by
the data of \citealt{San03} in their Fig.~5}. We find that the
total gas mass fraction within $\csim$, $\fxsim$, levels off to values
within about $\pm 10\%$ of the cosmic value for emission weighted
temperatures above 3--4 keV (i.e. for \emph{virial} masses above
(3--4)$\times 10^{14}$\msun), while it becomes directly proportional to
$\tx$ for less massive systems, which, in turn, implies that
$\fxsim\propto\msim^{0.6}$ in the hot group regime. This is a mass
dependence somewhat weaker than the self-similar behavior
$F_\mathrm{X}\propto M$ predicted by the models of \citet{BBP99} and
\citet{DD02}. Note also that our results are consistent with the
well-known fact that pressure forces arising from a high entropy floor
(see next section) can efficiently shut off the gas supply to the
halos, reducing the amount of baryons that end up gravitationally bound
\citep{CMT98,Mua02,OH03} and resulting therefore in smaller total
baryon fractions for halos with shallow potential wells.

\subsection{Scaling Relations for the Entropy}

A particularly interesting source of information about deviations from
self-similarity is the gas entropy, as it is a record of the
thermodynamic history of the diffuse baryons in clusters and groups of
galaxies. In X-ray studies this property is customarily represented
by the entropy \emph{of the gas electrons}, which we infer from the
expression
\begin{equation}\label{kx}
\kx(x)=\frac{\mub(\tstg,Z)^{5/3}\Vc^2}{(\zeta(\tstg,Z)\rhogc)^{2/3}}\frac{\tstg(x)}{\trhog(x)^{2/3}}\;,
\end{equation}
with $\zeta(T,Z)$ the function that renormalizes the number density of
gas particles to the number density of electrons\footnote{As in our
model $\gamma\neq 5/3$, the ratio $T/n_{\mathrm e}^{2/3}$ depends, for
a given $M$, on the halocentric distance.}. In the last few years, a
substantial amount of observations on the scaling properties for the
entropy on both the inner and outer regions of galaxy systems have
become available. As noted earlier in the Introduction, such
observations have served to establish that X-ray clusters and groups
show entropy excesses with respect to the expectations from pure
gravitational shock-heating \citep[e.g.,][]{PCN99,LPC00} that are not
restricted only to their central regions \citep{Fin02,PSF03}.

In order to confront our predictions with data from the
observational studies just mentioned, we have calculated the central
gas entropy (from eq.~[\ref{kx}]) at a scaled radius of $0.1\ctwohun$
against mean gas temperature within $0.3\ctwohun$, while to estimate
the entropy level on the halo outskirts we have chosen the adiabat at
$\csim$ scaled by $\msimfin^{-2/3}$, the total mass within this radial
distance expressed in units of $10^{13}$\msun, as a function of the
total system mass $\msim$.

Data and model predictions are compared in
Figure~\ref{adiabat}. For the inner entropy (left panel), our results
describe, reasonably well, the data points from \citet{PSF03} given
their large observational uncertainties. As shown in the plot, our
prediction deviates progressively from the $K\propto T$ scaling
followed by the high temperature systems, reaches a minimum value
between 1 and 2 keV, and rises again for the coldest
objects. Remarkably, this behavior is relatively consistent with the
trend described by the data in the pioneering studies of the entropy by
\citet{PCN99} and \citet{LPC00}, which predicted an entropy floor at
group scales with a typical value of 126\keVcm\ (for our adopted $h$)
not too different from the minimum value reached by our model, although
in their sparse data there was no sign of a recovery of the entropy at
the low-temperature end. However, the measurements in the much larger
sample gathered by \citet{PSF03} appear rather to delineate an 'entropy
ramp', meaning that the departure from the self-similar scaling is more
accurately represented by a gradual monotonic deviation characterized
by a logarithmic slope of about 2/3. Quite interestingly, though,
further data coming from elliptical galaxies \citep*{OPC03} ---which we
recall cannot be consistently explained by our modeling--- show some
objects with very high central entropies. Given the good agreement of
the predicted gas temperatures with observations (see
Fig.~\ref{m500tx}), this possible mismatch of the central entropy for
the coldest systems may be attributed to the fact that their associated
central gas densities are typically lower than actual measurements
\citetext{compare, for instance, our results in Sec.~\ref{profiles_3D}
with Fig.\ 8 of \citealt{SP03}}.

In Figure~\ref{adiabat} (\emph{right}) the predictions of our preferred
model for the scaling of the 'reduced' outer entropy,
$\kx(\csim)/\msimfin^{-2/3}$, against system mass are compared with the
data from \citet{Fin02}. As seen from the figure, in this case the mean
trend drawn by the full set of data points is followed with impressive
fidelity by the theoretical curve. Nevertheless, our results introduce
two major modifications with respect to the interpretation given by
Finoguenov and coworkers. First, we find that excess entropy should be
present across the full mass range of galaxy systems and not only in
those below a certain mass threshold \citetext{in line with the
interpretation given in \citealt{PSF03}}. And second, the scaled
entropy of the model shows a monotonic increase with decreasing halo
mass, thereby implying the absence of the 'entropy ceiling', i.e., of
the upper limit on the preheating, that was claimed to be present in
the \emph{ASCA} observations. As in \citet{Fin02}, we also have
included in the plot the entropy resulting from purely gravitational
heating (accomplished by setting $\DE$ to zero). Under these
conditions, we predict a nearly constant scaled entropy of $\sm
100$\keVcm\ ---consistent with the expectation that under
self-similarity conditions scaling by $\msim^{-2/3}$ should renormalize
the entropy to a value independent of system temperature--- that
clearly fails to explain the observations.

\subsection{Equilibrium Configuration of the Gas Distribution}\label{profs}

We now present our model predictions for the internal structure of the
IHM. We first focus on the spatially resolved properties of the
X-ray-emitting gas and then confront the gradients of the surface
brightness and projected temperature profiles with X-ray data.

\subsubsection{Intrinsic Profiles}\label{profiles_3D}

\begin{figure}
\centerline{\includegraphics*[width=9.5cm]{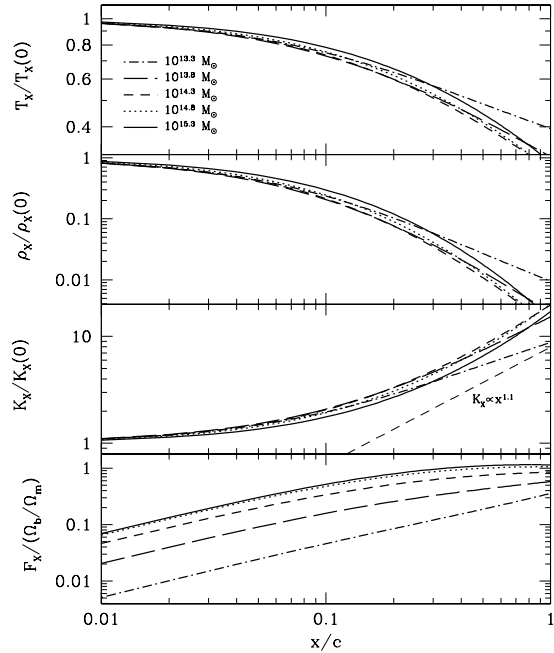}}
\figcaption{Predicted three-dimensional
radial profiles of the X-ray gas temperature, density, entropy, and
integrated gas mass fraction for five present-day halos of different
total masses. The three top panels show profiles scaled to their
central values, while the cumulative baryon fraction profile is given
in units of the cosmic value. Radial distances are normalized to
$\rvir$. The straight short-dashed line included in the third panel is
used to indicate the slope of 1.1 expected for the scaled entropy from
pure shock heating.
\label{profiles3D}}
\end{figure}

In Figure~\ref{profiles3D} we plot the three-dimensional radial
profiles of the scaled X-ray gas temperature, mass density, and
entropy, as well as of the integrated gas mass fraction, for a set of
representative halo masses at $z=0$. Radial distances are given in
units of $\rvir$ so as to facilitate the comparison among profiles of
different masses and with the results of previous works. Quite
remarkably, the gas temperature and density profiles (and hence the
entropy profiles) for halo masses above $6.3\times 10^{13}$\msun\ look
very similar when scaled to their central values, although we recall
that, in profiles of the NFW-like form, the radial coordinate must be
expressed in $\rs$ units in order to make it truly independent of the
halo mass (see, for instance, the model eqs.\ in \S~\ref{dark_matter}
and \ref{hotgas}). Only the less massive (coldest) systems tend to have
slightly shallower profiles as a result of the increasing influence of
preheating. In contrast, the bottom panel of Figure~\ref{profiles3D}
shows that the variation of the predicted gas fraction with radius,
which with the exception of the outer regions ($x\gtrsim 0.4c$) of the
most massive systems ($M\gtrsim 6.3\times 10^{14}$\msun) decreases
monotonically toward the halo centers, depends strongly on the halo
mass (temperature), mirroring the trend seen in Figure~\ref{fx500tx}
for the global values: the more massive the halos, the higher the
profile amplitude. This behavior agrees quite nicely with the most
recent observational determinations of this property by \citet{San03},
illustrating that the distribution of the IHM does not follow that of
the dark matter, but is significantly less concentrated as was expected
(see below).

The good agreement with observations is specially exciting for the
entropy profiles. On one hand, none of our profiles, even for the
smallest halos, is fully isentropic, thus invalidating one of the major
criticisms against universal preheating gas models consisting in the
production of flat entropy gradients for group-sized halos. On the
other hand, our model reproduces quite well, for all but the less
massive systems, the outer radii $K(r)\propto r^{1.1}$ behavior
observed by \citet{PSF03} and predicted in the gravitational
shock-dominated regime of entropy production by both theoretical
prescriptions based on spherical accretion within a NFW dark matter
halo \citep{ENF98,TN01,Bab02} and the latest state-of-the-art
cosmological gasdynamical simulations \citep{Bor04}. This result
indicates that it is possible to reconcile the large entropy excesses
seen at large radii, which suggest that entropy profiles are dominated
throughout by the effects of non-gravitational preheating, with the
fact that their typical outer slopes scatter about the value predicted
from shock heating, without requiring any specific fine tunning of the
model. Interestingly enough, the simulations by
\citet{Fin03} show that heating with an equal amount of energy per
particle produces entropy profiles that are similar to those arising
from accretion shocks of fixed strength.

A final aspect concerning the entropy profiles that we wish to point
out is that, if we apply the same scalings as in \citet{Fin02}, our
universal preheating model reproduces, very satisfactorily, the
behavior and amplitude at all radii of the deprojected profiles
measured by these authors \citep[see also][]{PSF03}. In particular, we
are capable of replicating the strongest rise with radius of the
richest clusters and both the flatter and higher entropy levels,
especially at smaller radii, found for groups. As we have not included
radiative cooling in our treatment, this suggests that the possible
effects of this latter mechanism might not be excessively important,
even in the central regions of massive clusters.

\begin{figure}
\centerline{\includegraphics*[width=9.5cm]{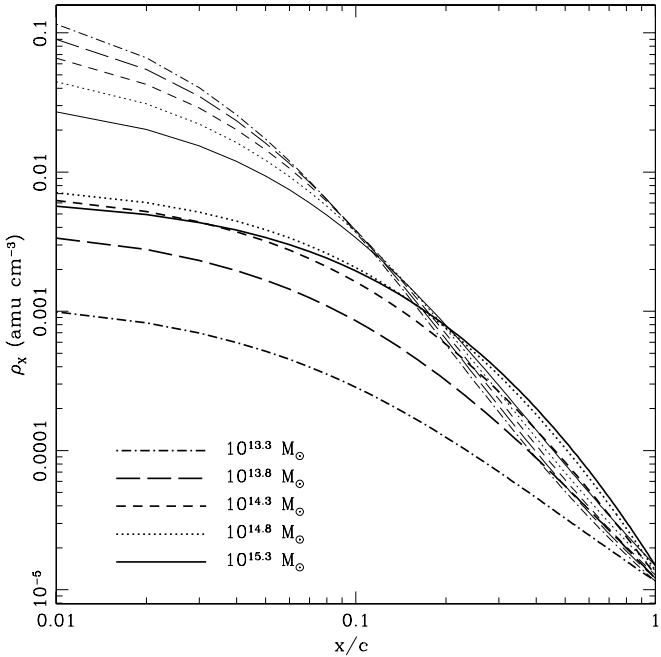}}
\figcaption{Comparison of the
predicted X-ray gas density profiles resulting from our preferred 
model (thick curves) and those resulting from a model with $\gamma=1$
and $\DE=0$ (thin curves). Each pair of equal-type curves
corresponds to a halo of a given current total mass. Radial distance is
in units of $\rvir$.
\label{gasdensity}}
\end{figure}

The three-dimensional radial profiles of the gas density have been also
redrawn in Figure~\ref{gasdensity} to provide an example of the impact
of preheating on the gas distribution. As in Figure~\ref{profiles3D},
we plot the profiles, now expressed in physical units to allow a direct
comparison with observations, corresponding to five halos with
different masses ranging from $2\times 10^{13}$ to $2\times
10^{15}$\msun. Two sets of curves are shown: one corresponding to our best
model and the other to the canonical $(\gamma,\DE)=(1,0)$
'non-radiative' conditions. It can be seen that preheating manifests
itself as a substantial reduction of the hot gas density in the central
halo regions. In agreement with expectations, this reduction is
strongest for the low-mass systems. Clearly, as the halo mass is
lowered, the gas becomes less concentrated and exhibits a bigger
core. Also evident is the fact that the gas distribution in the outer
halo regions for systems with $M\gtrsim 6.3\times 10^{13}$\msun\ is
more extended (i.e., the density at a given radius is raised) than that
corresponding to the $(\gamma,\DE)=(1,0)$ prediction. In contrast,
the effects of preheating lead to a substantial increase in the gas
temperature and entropy toward the halo centers \citetext{as similarly
reported by \citealt{MBB02} and \citealt{Bab02}}, not quite unlike the
predictions by galaxy formation-regulated gas evolution models
\citep[e.g.,][]{WX02}.

\subsubsection{Surface Brightness Profiles}

The observed X-ray surface brightness at a \emph{projected} scaled (by
$\rs$) radius $X$ from the center of a spherical halo is given simply
by the Abel integral\footnote{While strictly the integral along the
line-of-sight (\ref{Sx}) should extend to infinity, the applicability
of the Jeans equation (\ref{Jeans}) beyond $\rvir$ becomes
questionable.}
\begin{equation}\label{Sx}
S_X(X)=2\rs\int^{\sqrt{c^2-X^2}}_{0}\epsilon_{\Delta\nu}(x)dl\;,
\end{equation}
where $l=\sqrt{x^2-X^2}$ is the distance along the line-of-sight and
$\epsilon_{\Delta\nu}(x)$ is the X-ray volume emissivity within the
energy band $\Delta\nu$ of the observations. It turns out that our
predicted X-ray surface brightness profiles \emph{in the} ROSAT
\emph{broad (0.1--2.4 keV) band} are a reasonable match to the
conventional $\beta$-model,
$S_X(R)/S_X(0)=[1+(R/\rc)^2]^{-3\beta+1/2}$, within $\csim$. Beyond
this radius, the projected gas emission initiates a rapid downfall due
to the finite outer boundary of the integral (\ref{Sx}). Thus, for the
purposes of comparison with the existing X-ray imaging measurements,
the resulting $S_X$ profiles of present-day halos have
been matched by $\beta$-models over the radial interval
$0.01\csim<X<\csim$, which corresponds approximately to extending the
fit out to half the virial radius, and the best-fit values of the
slope $\beta$ and core radius $\rc$ have been compared with their
observational determinations. We note that the portion of the profile
being fitted is consistent with the fact that measurements of the X-ray
surface brightness of clusters and groups generally embrace only a
fraction of their total radius.

\begin{figure}
\centerline{\includegraphics*[width=12.5cm]{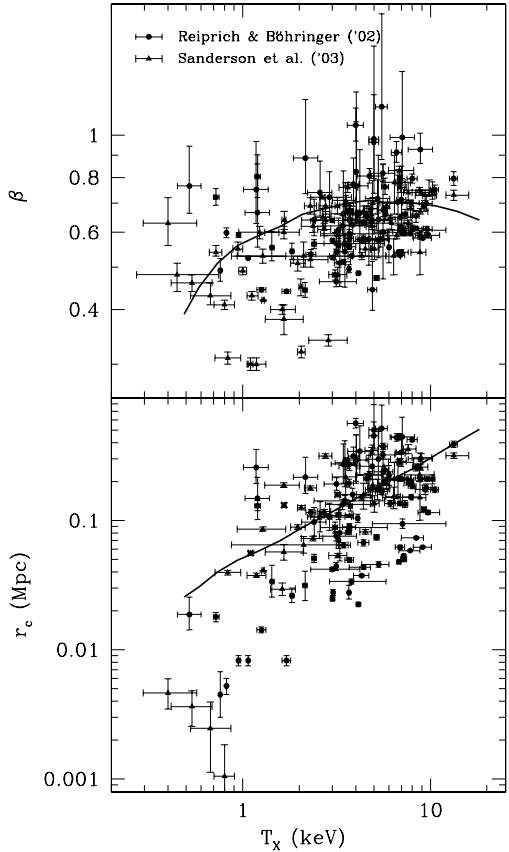}}
\figcaption{Temperature dependence of
the slope parameter (\emph{top}) and core radius (\emph{bottom})
inferred from $\beta$-model fits to the X-ray surface brightness
profiles. Different datasets are identified with different
symbols. Curves show the results of fitting within the radial range
$0.01\csim<X<\csim$ the $S_X$ in the 0.1--2.4 keV band predicted by our
preferred model.
\label{betarc}}
\end{figure}

Figure~\ref{betarc} shows how the predicted dependence of our
best-fit $\beta$-model parameters on $\tx$ compares to data
extracted from the same catalogs used in the investigation of the
mass-temperature relation, with the emission-weighted temperatures
quoted by \citet{San03} conveniently rescaled (see \S~\ref{msimtx}).
Although the large dispersion in each parameter at a given temperature
does little to constrain the model predictions, our theoretical curves
fare, once more, quite satisfactorily, as they appear to roughly track
the mean trends that would be inferred from 'chi-by-eye' fits to the data
points. Quite remarkably, the prediction for the slope parameter
supports the frequently advocated case for flatter surface brightness
profiles for groups \citep[e.g.,][]{PCN99,HP00,Mul03,OP04}, while for
high-mass systems it typically approaches the self-similar value of 2/3
\citep{JF84}. On the other hand, the predicted increase of the central
gas density concentration with the $\sm 3/4$ power of the gas
temperature is also consistent with the data, which suggests further a
possible sharp drop of $\rc$ for systems with X-ray temperatures
$\lesssim$\,1 keV. Regarding this latter possibility, we point out
that, as shown by \citet{Voi02}, observational systematics such as the
range of radii in the fit and surface brightness bias can affect
measurements for low temperature systems. In the former case, we have
verified that fitting over the radial range $0.01\csim<X<0.5\csim$
essentially involves an overall reduction of the amplitude of the
predictions, while raising the surface brightness threshold, which
makes the best-fit values of $\beta$ and $\rc$ both decline faster
with decreasing $\tx$ has, however, the undesirable consequence of
excluding from detection the coldest groups whose distinct behavior
one is trying to describe.

\subsubsection{Projected Temperature Profiles}

\begin{figure}
\centerline{\includegraphics*[width=9.5cm]{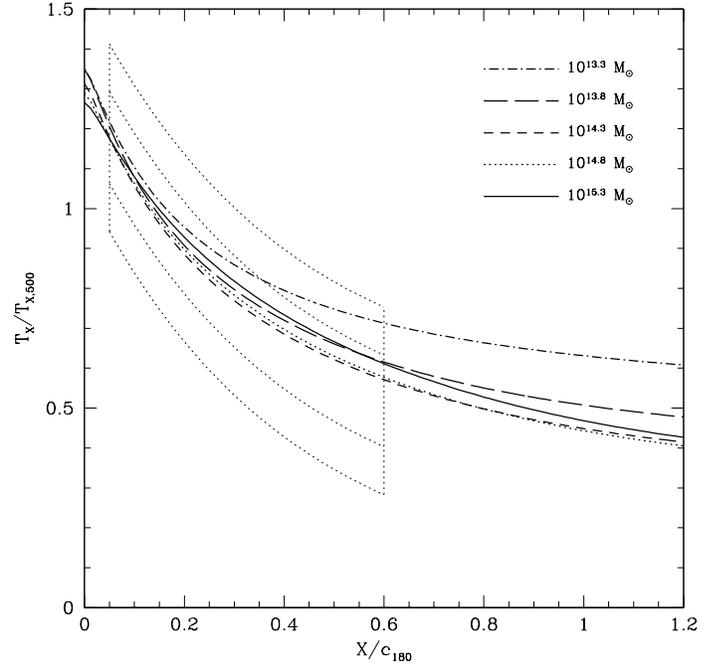}}
\figcaption{Predicted projected emission-weighted X-ray gas temperature
profiles resulting from our best model, scaled by $\txsim$. Thick
curves represent five halos of different total masses. The dotted boxes
enclose the \emph{ASCA} temperature profiles observed by \citet{Mar98}:
the outer one shows the $90\%$-error band of all their temperature
profiles, while the inner box approximates the scatter of the
corresponding best-fit polytropic models. Average temperatures of the
observed clusters range from 3 to 10 keV, which in our model correspond
to total halo masses ranging from $\sm 2\times 10^{14}$ to $2\times
10^{15}$\msun. To facilitate comparison with data projected radial
distance is in units of $r_\mathrm{180}$.
\label{projtx}}
\end{figure}

As a last check of the consistency of our model, we have calculated
the two-dimensional projection of the emissivity-weighted temperature
profile of the halos (eq.~[\ref{tx}]), which is defined as

\begin{equation}\label{tx2D}
\tx(X)=\frac{\Vc^2\int^{\sqrt{c^2-X^2}}_{0}\mub(\tstg,Z)\tstg(x)\epsilon(x)dl}{\int^{\sqrt{c^2-X^2}}_{0}\epsilon(x)dl}\;.
\end{equation}

Figure~\ref{projtx} illustrates the predictions by our preferred model
for the projected emission-weighted temperature profiles of five halos
with present-day masses equally spaced (in logarithmic scale) between
$2\times 10^{13}$ and $2\times 10^{15}$\msun, which rise uninterrupted
toward the halo centers. Interestingly, our results for masses above
$2\times 10^{14}$\msun\ (i.e., $\tx\gtrsim 3$ keV) compare very
favorably with the composite radial temperature profile by
\citet{Mar98}. From an analysis of \emph{ASCA} spatially resolved
spectroscopic data for a set of nearby clusters with mean spectral
temperatures above 3 keV, these authors found that the radial run of
the temperature for symmetric, cooling-flow corrected, clusters showed
a nearly universal outward decline between $\sm 0.1$ and $\sm
0.6\coneight$ corresponding to a polytropic index
$1.24^{+0.20}_{-0.12}$ (with $90\%$ errors). We note that the model
profiles have been scaled by $\txsim$, the average temperature within
$\csim$, in an attempt to represent more accurately the scaling by the
average temperatures calculated over the radial range covered by
\citeauthor{Mar98}'s data. A different (similarly realistic)
normalization would have only a marginal impact on the profiles and
would not change our conclusions.

Albeit in another work based on \emph{ASCA} observations \citet{FAD01}
reported a similarly outward decreasing behavior, the results by
\citet{Mar98} have nonetheless been controversial, with some other
\emph{ASCA} papers claiming consistency with overall isothermality
\citep{Whi00}. Besides, there is the well-known analysis of
\emph{Beppo-SAX} data by \citet{DM02}, who derived profiles similar to
those of \citet{Mar98} in the outer cluster regions, but not in the
very centers, where they observed flat temperature gradients for radii
smaller than $\sm 0.2\coneight$. Other \emph{Beppo-SAX} studies of a
more limited radial extent produced temperature profiles that were
generally flat or even rising slightly out to $\sm 30\%$ of the virial
radius \citep{IB00}. Now, however, the latest results by \emph{Chandra}
appear to confirm quite nicely the steady decrease of the gas
temperature in the $0.1-0.6\coneight$ radial range, although the
central peak at $x < 0.1\coneight$ looks like an artifact of the
cooling flow modeling, as \emph{ASCA} could not resolve the centers
well\footnote{As our model, by design, ignores radiative cooling, we
cannot account for the smooth decline of the temperature toward the
center observed in the innermost regions ($\lesssim 0.1\coneight$) of
some massive clusters with strong cooling flows
\citep*[e.g.,][]{ASF01}.} (Markevitch 2004, private
communication). Further support to an outward decreasing IHM
temperature over a great deal of the radial range comes from the
results of gasdynamic simulations both adiabatic \citep[e.g.,][]{Asc03}
and with cooling, star formation, and supernova feedback
\citep{Bor04,Ett04}.

\section{SUMMARY AND CONCLUSIONS}

We have investigated the global and structural X-ray properties of the
IHM in nearby groups and clusters of galaxies within a flat
$\Lambda$CDM cosmology using a simple, reasonably well specified
analytic polytropic gas model relying on the inside-out growth of
structure between major mergers. Among the different scenarios
suggested in the literature as a means of explaining the
well-established breakdown from the simplest self-similar predictions,
we have focused on the one which assumes that the specific energy of
the gas is increased by some arbitrary source of extragravitational
heating before the virialization of galaxy systems. This additional
injection of energy onto the intergalactic medium has been presumed to
be independent of the virial mass of the halo where the hot gas is
incorporated. We summarize our main findings below.

(i) A polytropic index $\gamma=1.2$, as many observations and numerical
simulations suggest, offers the best consistency with the postulate
that the specific energy balance between the two main halo components
(dark matter and hot gas) is preserved when energy loses are
negligible. This is the first time a theoretical model tackling the
X-ray properties of relaxed galaxy systems ---the condition that the
preheated gas remains bound constraints our predictions to halos with
present-day total virial masses above $\sm 2\times 10^{13}$\msun---
provides a possible physical origin for such a value of $\gamma$. A
natural consequence of this hypothesis is that the IHM properties must
be essentially independent of the system formation redshift. Variations
on the metallicity of the hot gas could be a possible alternative to
claims that favor differences in the formation epoch as the primary
source of the dispersion in the observed scaling relations at group
($\tx\lesssim 2$ keV) scales.

(ii) With $\gamma$ fixed to 1.2, our hot X-ray gas model for group- and
cluster-sized halos is capable of describing very accurately the
observed mean trends of a set of representative scaling laws between
bulk properties of this constituent for a level of energy injection of
0.55 keV per gas particle, low enough to be accessible to popular
nongravitational heating sources such as supernovae and AGN. In
particular, we infer mass-temperature relationships with slopes steeper
than the self-similar prediction. For the $\msim-\tx$ relationship we
find a nearly constant logarithmic slope of $\sm 1.7$ across most of
the temperature range that initiates a slight offset toward lower
masses for $\tx\lesssim 1$ keV pointing to the locus occupied by
early-type galaxy data. On the other hand, the $\mxsim-\tx$ correlation
exhibits a slope similar to the $\msim-\tx$ relation for the hottest
clusters that progressively steepens as the scale diminishes,
suggesting that for group-sized systems the total gas fraction within
the halo should grow steadily with mass approximately as $M^{0.6}$. The
model also allows us to confidently rule out purely isothermal
($\gamma=1$) or adiabatic ($\gamma=5/3$) descriptions of the X-ray
data.

(iii) The scaling properties of the gas entropy are similarly well
predicted. According to the results of our investigation, a preheating
stage involving a relatively modest amount of energy injection of about
half a keV per particle, suffices to explain, simultaneously, the near
100\keVcm\ level shown by the central entropy of halos at group scales
and the higher entropy level of $\sm 400$--500\keVcm\ seen in the
periphery of these galaxy systems.

(iv) Only the prognosticated behavior of the inner entropy for the
coldest systems appears to be in conflict with
observations. Specifically, in halos with emission-weighted
temperatures below $\sm 1$ keV our model indicates that the central
entropy should increase with decreasing gas temperature, probably due
to the substantial reduction in their central gas densities caused by
preheating. In contrast, the preferred interpretation of the most
recent observational data is that the central entropy obeys a
non-self-similar power law form, $\kx\propto\tx^{2/3}$, at all
scales. However, the systematic and statistical uncertainties in
observations resulting from limitations in the instrumental response
and the intrinsic faintness of the groups are sufficiently important
that a strong discrepancy cannot be claimed. Alternatively, one may
also consider the possibility that a universal preheating model does
not provide an accurate description of the IHM physics in the innermost
regions of cold galaxy systems. This would be the case if the epoch of
preheating occurs relatively late in the history of the universe, say
at $z\sm 2$, so that most of the group cores, which tend to form early
($\zf\sm 3$--4) and settle later in the central regions of larger
units, would be already in place when the energy boost takes place
\citetext{calculations by \citealt{OB03} using the Press-Schechter
theory show that at $z=2.2$ about $50\%$ of the cluster progenitors
would not be affected by preheating}. This implies that heating could
have been less efficient in raising the entropy of less massive systems
because of the high density of their IHM, whose thermodynamic history
would be then dominated by the small amounts of entropy acquired
through gravitational shocks.

(v) The equilibrium gas profiles also agree quite nicely with recent
observational results, suggesting that the assumed inside-out growth of
structure after virialization offers a straightforward and sensible
approximation to the evolution of the gas equilibrium profiles. Of
particular relevance are our findings that the IHM does not show,
contrarily to other pure preheating models, a largely isentropic core
and that the outer logarithmic slopes of the entropy profiles for all
but the coldest systems are quite similar to the 1.1 value expected
from shock heating ---without necessarily implying that this is the
basic mechanism responsible for generating the observed excess entropy
at large radii in groups and clusters of galaxies.

(vi) We infer, in addition, outward declining temperature profiles,
much like most of the observational studies report outside the cluster
centers and in line with the trends found in recent hydrodynamical
simulations. Claims that the hot gas distribution within cluster-sized
dark halos is well represented by a polytropic model with $\gamma\sm
1.2$ are reinforced further by the behavior predicted for the radial
distribution of the projected emission-weighted temperature in
relatively hot clusters (i.e., $\tx\gtrsim 3$ keV). However, our model
cannot account for either the nearly isothermal profiles or the large
flat cores detected in certain recent observational analyses of
temperature gradients in clusters. Although the failure to reproduce
these trends could reflect the need for incorporating some additional
physics, such as a heat transport mechanism (e.g., thermal conduction)
capable of bringing large regions of the IHM to the same temperature in
a relatively short timescale, the fact is that the actual form of the
profile that sets the norm for bound galaxy systems has yet to be settled.

Our conclusion from these results is that, in general terms, our
universal preheating model predictions are very succesful in fitting
the observational data. The good match is more remarkable if one takes
into account that it has been achieved without resorting to elaborated
schemes of energy injection and the additional freedom of including the
effects of observational biases in the measured gas
properties. Admittedly, processes deliberately excluded from our
modeling, such as the condensation and removal of the lowest entropy
gas from the IHM, must happen at some level because that is how
galaxies and stars form within dark halos. Moreover, radiative cooling
can help to reduce the energetic requirements for preheating while it
might be necessary in order to explain the behavior of the temperature
profiles in the innermost regions ($r\lesssim 0.1\rvir$) of massive
cooling-flow clusters revealed by new high-spatial resolution X-ray
data. Nevertheless, the fact that the picture we are proposing can
account, within current observational constraints, for the typical
X-ray characteristics of nearby galaxy systems argues in favor of it as
a useful description of the evolution of the intergalactic medium
properties in the densest regions of the universe.

Yet the most important asset of the present work is perhaps that
the X-ray properties of galaxy systems ranging from the coolest groups
to the hottest clusters have been intimately linked to a successful
analytic theory of the clustering history in the universe, thus
providing an intertwined treatment for the formation and evolution of
the structure of both nonbaryonic and baryonic matter that can be
easily implemented in models of galaxy formation.

\begin{acknowledgements}
We thank Stefano Borgani, Alexis Finoguenov, Maxim Markevitch, Alastair
Sanderson, and Thomas Reiprich for kindly providing the data shown in
some of the figures. This work was supported by the Direcci\'on
General de Investigaci\'on Cient\'{\i}fica y T\'ecnica of Spain, under
contract AYA2003--07468--C03--01.
\end{acknowledgements}

\singlespace

\end{document}